\documentclass[prd,twocolumn,showpacs,nofootinbib]{revtex4}

\usepackage{times}
\usepackage{natbib}
\usepackage{epsfig}

\newcommand{\DS}{\Delta\Sigma}
\newcommand{\rms}{\sigma_8}
\newcommand{\MB}{\bar M} 
\newcommand{\Mmin}{M_\up{min}}
\newcommand{\Mmax}{M_\up{max}}
\newcommand{\msun}{M_\odot}
\newcommand{\hmsun}{{h^{-1}\msun}}
\newcommand{\up}[1]{{\rm #1}}
\newcommand{\mpc}{{\rm Mpc}}
\newcommand{\hmpc}{{h^{-1}\mpc}}
\newcommand{\OM}{\Omega_m}
\newcommand{\beeq}{\begin{equation}}
\newcommand{\kmin}{k_\up{min}}
\newcommand{\kmax}{k_\up{max}}
\newcommand{\eneq}{\end{equation}}
\newcommand{\plin}{P_\up{lin}}
\newcommand{\XX}{\mathcal{C}}      
\newcommand{\hmpci}{{h\mpc^{-1}}}
\newcommand{\pvec}{\bdv{p}}
\newcommand{\nbg}{n_{bg}}          
\newcommand{\ZZ}{\omega}           
\newcommand{\YY}{\mathcal{G}}      
\newcommand{\sigM}{\sigma_\up{ln M}}
\newcommand{\NN}{n_g}              
\newcommand{\bdv}[1]{{\bf #1}}
\newcommand{\OO}{O}      
\newcommand{\erfc}{\rm{erfc}}
\newcommand{\bear}{\begin{eqnarray}}
\newcommand{\enar}{\end{eqnarray}}
\newcommand{\AVE}[1]{\langle#1\rangle}

\begin{document}

\title{Joint Analysis of Gravitational Lensing, Clustering and Abundance:\\
Toward the Unification of Large-Scale Structure Analysis}

\author{Jaiyul Yoo$^{1,2}$}
\altaffiliation{jyoo@physik.uzh.ch}
\author{Uro{\v s} Seljak$^{1,2,3,4}$}
\affiliation{$^1$Institute for Theoretical Physics, University of Z\"urich,
CH-8057 Z\"urich, Switzerland}
\affiliation{$^2$Lawrence Berkeley National Laboratory, University of 
California, Berkeley, CA 94720, USA}
\affiliation{$^3$Physics Department and Astronomy Department,
University of California, Berkeley, CA 94720, USA}
\affiliation{$^4$Institute for the Early Universe, Ewha Womans University, 
120-750 Seoul, South Korea}

\begin{abstract}
We explore three different methods based on weak lensing
to extract cosmological constraints from the large-scale structure. 
In the first approach (method~I: Seljak {\it et. al.} 2005),
small-scale galaxy or cluster lensing measurements of their halo mass provide
a constraint on the halo bias, which can 
be combined with the large-scale galaxy or cluster clustering 
to measure the dark matter clustering. 
In the second approach (method~II: Baldauf {\it et. al.} 2010), 
large-scale galaxy clustering and large-scale galaxy-galaxy lensing 
each trace the large-scale dark matter clustering, and 
the two can be combined into a direct 
measurement of the dark matter clustering. 
These two methods can be combined into one method I+II
to make use of lensing measurements on 
all scales.
In the third approach (method~III), we add abundance information to 
the method~I,
which is a version of self-calibrated cluster abundance method. 
We explore the statistical power of these three approaches as a function of 
galaxy or cluster luminosity to investigate
the optimal mass range for each method and their cosmological constraining 
power. In the case of the SDSS, we find that the three methods give 
comparable constraints, but not
in the same mass range: the method~II works best 
for halos of $M\sim10^{13}\msun$,
typical of luminous red galaxies, and the methods~I and~III work best for 
halos of $M\sim10^{14}\msun$, typical of low mass clusters. 
We discuss the robustness of each method against various systematics.
Furthermore, we extend the analysis to the future large-scale galaxy surveys
and find that the cluster abundance method is {\it not} 
superior to the combined method I+II,
both in terms of statistical power and robustness against systematic errors.
The cosmic shear-shear correlation analysis in the future surveys yields 
constraints as strong 
as the combined method, but suffer from additional systematic 
effects. We thus advocate the combined analysis of clustering and lensing 
(method I+II) as a powerful alternative to other large-scale probes. 
Our analysis provides a guidance to
observers planning large-scale galaxy surveys such as the DES, Euclid,
and the LSST. 
\end{abstract}


\maketitle

\section{Introduction}
Large-area galaxy surveys like the Sloan Digital Sky Survey (SDSS; 
\cite{YOADET00,SDSSDR7})
have enabled high-precision measurements of galaxy
clustering over a wide range of separation, and galaxy clustering has now
become a commonly exercised and indispensable tool in cosmology.
While galaxies are generally expected to trace the dark matter distribution 
up to an overall factor on large scales \cite{KAISE84}, its relation becomes
complicated on small scales, and inferring cosmological parameters
from galaxy clustering measurements are, therefore, hampered by the
complex relation between the galaxy and the dark matter distributions,
known as galaxy bias $b_g$.
In the linear regime, measurements of the galaxy power spectrum in redshift
space \cite{KAISE87} provide ways to break the degeneracy by constraining 
the parameter combination $f/b_g$, where $f$ is the logarithmic growth rate.
However, the recent analysis of the SDSS redshift-space distortion yields
a relatively large uncertainties $\sim30\%$ in the linear regime
\cite{TESTET04},
and its full potential remains to be realized in future surveys with
larger sky coverage (see, e.g., \cite{SELJA01,WHITE01,TWZ06,SEMC11,REWH11}
for other approaches to modeling redshift-space distortions on small scales).

Gravitational lensing uses the subtle distortion of background source galaxy
shapes to statistically map the foreground matter distribution that
causes the gravitational shear (e.g., \cite{MELL99,BASC01,REFRE03}).
Especially, galaxy-galaxy lensing
measures the distortion in background source galaxy shapes around the
foreground lensing galaxies \cite{TYVAET84,BBS96,FIMCET00,SHJO04,MSKHB06}.
With the statistical power present
in the SDSS, high precision ($20-30\sigma$) measurements of galaxy-galaxy
lensing signals are typically available for various galaxy samples, making it 
a useful tool for cosmology (see, e.g., 
\cite{MSKHB06,MSCBHB06,MASEET10}). Furthermore,
spectroscopic redshift measurements of the foreground lens galaxies
allow the galaxy-galaxy lensing signals $\gamma$ 
to be related to the excess surface
density $\DS$ at the lens redshift \cite{MIRA91}, which in turn is related
to the galaxy-matter cross-correlation $\xi_{gm}$.

The combination of galaxy-galaxy lensing and galaxy clustering measurements
is, therefore, helpful in breaking the degeneracy in galaxy bias~$b_g$ 
and measuring the matter fluctuation amplitude~$\rms$ (see, e.g., 
\cite{GUSE02,HUJA04,MTSKW05,
YTWZKD06,CAVAET09,BASMET10,LETIET11,CAVAET12}).
In this work we perform a systematic investigation of the cosmological
constraining power that can be derived in the current and future galaxy
surveys by combining both measurements of galaxy clustering and galaxy-galaxy
lensing. In particular, we are interested in gaining the physical insights
of the resulting constraints in a model independent way.

\begin{table*}
\caption{SDSS galaxy samples. The approximate numbers for the SDSS galaxy
samples are taken \cite{ZEZHET11,MASEHI08} to represent
the SDSS Main, LRG, and maxBCG samples. The mean mass $\MB$
of the galaxy samples
is obtained from the galaxy-galaxy lensing measurements 
\cite{MSKHB06,MASEHI08}. The minimum mass $\Mmin$ and the maximum mass
$\Mmax$ are obtained by matching the number density and the mean mass of
each sample. All masses are in units of $\hmsun$.}
\begin{tabular}{ccccrccccccccccc}
\hline\hline
sample & & $M_r^{0.1}+5\log h$ & &$N_\up{tot}$& & $n_g~(\hmpc)^3$ & & $\MB$ && 
$\Mmin$ && $\Mmax$ && $\bar z$ \\
\hline
L1 &&$-18$ to $-17$  && 5900   && 2.0$\times10^{-2}$ && 8.8$\times10^{10}$ && 
$6.1\times10^{10}$ && $1.3\times10^{11}$ && 0.03\\
L2 && $-19$ to $-18$ && 18,000 && 1.3$\times10^{-2}$ && 3.4$\times10^{11}$ && 
$1.7\times10^{11}$ && $8.2\times10^{11}$ && 0.04\\
L3 && $-20$ to $-19$ && 44,000 && 1.0$\times10^{-2}$ && 4.3$\times10^{11}$ && 
$2.2\times10^{11}$ && $9.5\times10^{11}$ && 0.06\\
L4 && $-21$ to $-20$ && 100,000&& 5.3$\times10^{-3}$ && 1.2$\times10^{12}$ && 
$5.2\times10^{11}$ && $3.6\times10^{12}$ && 0.10\\
L5 && $-22$ to $-21$ && 69,000 && 1.0$\times10^{-3}$ && 5.4$\times10^{12}$ && 
$2.9\times10^{12}$ && $1.2\times10^{13}$ && 0.15\\
LRG && $-23.6$ to $-21.6$ && 62,000 && 1.0$\times10^{-4}$ && 3.8$\times10^{13}$ && 
$2.3\times10^{13}$ && $7.4\times10^{13}$ && 0.28\\
BCG1 && $-24.0$ to $-22.5$ && 8500   && 3.0$\times10^{-5}$ && 1.0$\times10^{14}$ && 
$6.1\times10^{13}$ && $2.1\times10^{14}$ && 0.25\\
BCG2 && $-24.0$ to $-22.5$ &&  850   && 3.0$\times10^{-6}$ && 3.0$\times10^{14}$ && 
$2.1\times10^{14}$ && $5.8\times10^{14}$ && 0.25\\
BCG3 && $-24.0$ to $-22.5$ &&   85   && 3.0$\times10^{-7}$ && 6.0$\times10^{14}$ && 
$4.7\times10^{14}$ && $2.8\times10^{15}$ && 0.25\\
\hline
\end{tabular}
\label{tab:sdss}
\end{table*}

In response to the recent development in numerical simulations and 
large-scale galaxy surveys, a halo model based approach to modeling galaxy
clustering and galaxy-galaxy lensing has been developed 
\cite{SELJA00,MAFR00,PESM00,SCSHET01,BEWE02,GUSE02}.
The key part of these approaches is to assume the halo occupation distribution
(HOD) and the spatial distribution of satellite galaxies, and to relate the
dark matter distribution to the galaxy distributions (see, e.g., 
\cite{COSH02}).
However, many parameters and assumptions of the models make it somewhat 
difficult to untangle the true cosmological constraining power. 
In particular, small-scale galaxy clustering information (roughly defined to 
be below 
twice the virial radius of the largest halos) is used for inferring 
HOD parameters of the model such as the satellite fraction and their 
radial distribution inside halos.
While this information provides useful constraints on HOD parameters, 
it is difficult to derive any useful cosmological information with it. 
Moreover, the limited number of HOD 
parameters explored to date may artificially provide tighter constraints on
the cosmological parameters. To avoid this concern we take a
simpler approach to the problem by approximating galaxy samples as 
individual halos with a certain range of mass. This approach is equivalent to 
identifying central galaxies and to removing satellites in the galaxy samples.
Therefore, there is no useful galaxy clustering signal on small scales.

Without the complication of the spatial galaxy distribution and the
halo occupation distribution, we can compute the galaxy clustering and 
gravitational lensing signals in a robust and model-independent way.
Galaxy clustering arises from the halo clustering and is tracing a biased 
version of the dark matter clustering on large scales,
while galaxy-galaxy lensing can be split into two regimes. 
Small-scale galaxy-galaxy lensing measures the density profiles of
dark matter halos and estimates its mass, and large-scale galaxy-galaxy
lensing measures the cross-correlation of the dark matter and the halo
distributions.
We analyze three different methods to extract cosmological information. 
In the method~I, we use small-scale lensing around galaxies to determine 
the halo mass, which in turn determines their large-scale
bias using theoretical mass-bias relation. 
Once the bias is known one can use galaxy auto-correlation to determine the dark matter 
clustering. This approach was first attempted in \cite{SEMAET05} using the
Main sample of the SDSS galaxies 
with limited success. We will show here that higher mass samples offer a better chance of 
success. 
The method~II combines large-scale lensing and clustering to eliminate bias, and 
has been developed in detail in \cite{BASMET10}. The method~III adds abundance information 
to the method~I, and becomes a specific implementation of 
the cluster abundance method \cite{HU03,LIHU04}.
Traditional cluster abundance methods 
rank clusters by their mass and determine their abundance as a function of it. 
Two main issues in cluster abundance methods 
are getting correct mass of the clusters and determining 
the scatter between the mass observable and the mass, since both mass calibration and scatter
are completely degenerate with cosmological parameters \cite{MASE07}. 
One can determine mass calibration with small-scale lensing 
and scatter with large-scale 
clustering analysis, hence the method~III uses the same 
information as the method~I with added cluster abundance information. 

In this paper, we investigate the cosmological constraining
power of combining galaxy clustering and galaxy-galaxy lensing and focus
on finding galaxy samples that are best suited for this purpose in the SDSS.
Furthermore, we attempt to answer the same questions in future galaxy surveys
by extending our analysis to higher redshift. Since numerous ongoing and 
planned future surveys are equipped with deep and wide imaging capability,
there exists another and potentially more powerful way to avoid the
complication of galaxy bias and to directly map the matter distribution:
Using the
cosmic shear-shear power spectrum \cite{BSBV91,MIRA291,KAISE92}.
Measurements of the
auto-correlation of subtle shape distortions, however, have proved to be
difficult due to the intrinsically weak signal-to-noise ratio and numerous
systematic uncertainties \cite{HIMAET04,MHSGET05,HUTAET06,BLMCSE11,NAMAET12}.
Since the first detections 
\cite{BAREEL00,KAWILU00,WITYET00,VAMEET00}, only a handful of
new measurements have appeared,
mostly using narrow but deeply imaged areas, such as 
from the Hubble Space Telescope \cite{HEBRET05,LEMAET07},
and the Canada-France-Hawaii Telescope \cite{HOMEET06}.
Repeat imaging of the narrow stripe-82 in the SDSS
enables measurements of cosmic shear of about 200 square degrees \cite{HUHIET11,HUEIET11,LIDOET11},
and no larger-scale weak lensing surveys exist yet. We consider the cosmic shear
measurements as another component of gravitational lensing 
in future galaxy surveys and compare the resulting constraints to the
combined constraints of galaxy clustering and galaxy-galaxy lensing.
In our analysis we do not include redshift-space distortions, which is another 
way to extract the information about the galaxy bias. Redshift-space 
disortions suffer from significant nonlinear and scale-dependent bias issues 
\cite{OKSEDE12} and need to be 
understood better before they can be used for high precision cosmology. 

The organization of the paper is as follows. 
In Sec.~\ref{sec:samples} we present our model for the SDSS galaxy samples
and discuss their physical properties. In Sec.~\ref{sec:combine}
we present three model-independent ways to combine measurements of 
gravitational lensing and galaxy clustering: Small-scale galaxy-galaxy
lensing and large-scale galaxy clustering in Sec.~\ref{ssec:metI}, 
large-scale galaxy-galaxy lensing and large-scale galaxy clustering in
Sec.~\ref{ssec:metII}, and additional abundance information in 
Sec.~\ref{ssec:abun}. Large-scale galaxy clustering is briefly discussed
in Sec.~\ref{ssec:lss}, and the combination of various methods is presented in
Sec.~\ref{ssec:full}. In Sec.~\ref{sec:fut}, we extend our analysis to the
future galaxy surveys such as the DES, BigBOSS, Euclid, and the LSST.
The cosmic shear constraints are complemented and compared to the galaxy-galaxy
lensing and galaxy clustering constraints in the future surveys.
We summarize our results and discuss the implications for planning large-scale
galaxy surveys in Sec.~\ref{sec:discussion}. Our calculations are performed
by assuming a flat $\Lambda$CDM universe with the matter density
$\OM=0.23$, the matter fluctuation amplitude $\rms=0.81$, and the spectral
index $n_s=0.97$. The matter power spectrum shape
is kept fixed in cosmological parameter variations.

\begin{figure*}
\centerline{\psfig{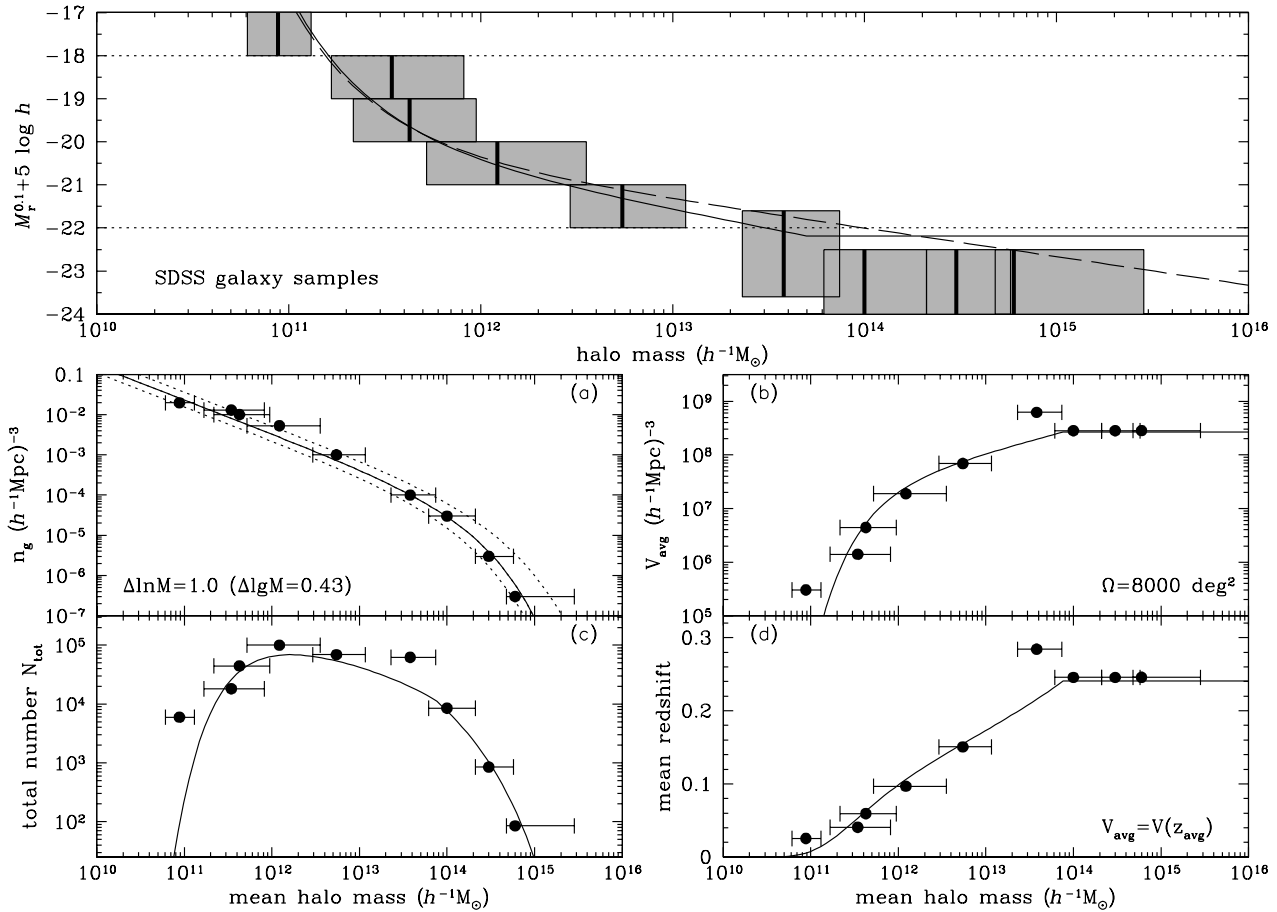}}
\caption{SDSS galaxy  and continuous mass-bin samples. We approximate
galaxy samples as isolated halos and use the relation between 
central galaxy luminosity and halo mass to match the observed SDSS galaxy
samples (discrete points, see Table~\ref{tab:sdss}). 
Top panel: Rest-frame $r$-band magnitude at $z=0.1$. Vertical solid lines
show the mean mass of the SDSS galaxy samples obtained from the galaxy-galaxy
lensing measurements \cite{MSKHB06,MASEHI08}. The minimum and the maximum
masses represented by gray boxes are obtained by matching the mean mass and
the number density of the SDSS galaxy samples. The dashed curve shows the
mass-luminosity relation \cite{ZEZHET11} between the central galaxy luminosity
and its halo mass, obtained by analyzing the clustering
measurements of the SDSS Main galaxy samples. Its validity is limited
to $-22\leq M_r^{0.1}+5\log h\leq-18$ (horizontal lines).
We modify the mass-luminosity relation shown as the solid curve
to implement the LRG and the maxBCG
samples at high luminosity, but account for the flattening of the
luminosity of central galaxies at $M>5\times10^{13}\hmsun$.
Bottom panels: The average volume and
the mean redshift of the SDSS galaxy samples are obtained by dividing the
observed number density by the total number of observed galaxies. For the 
continuous mass-bin samples (solid), we adopt the mass-bin interval 
$\Delta\ln M$ (dotted) that closely
matches the observed number densities of the SDSS galaxy samples, and we 
compute the average volume and the mean redshift by using the mass-luminosity
relation.}
\label{fig:samples}
\end{figure*}

\section{SDSS Galaxy Samples}
\label{sec:samples}
Here we describe our simple model for the SDSS galaxy samples, which will
be used to investigate their cosmological constraining power.
The SDSS \cite{YOADET00,SDSSDR7}
completed its observation in 2008 and mapped the sky
over 8000 deg$^2$ in five photometric bands to a limiting magnitude 
$r=$22.5. We consider nine galaxy samples that closely match the observed
SDSS galaxy luminosity-bin samples:
The SDSS Main galaxy (L1$\sim$L6), the luminous red galaxy (LRG),
and the maxBCG samples (e.g., 
\cite{ZEEIET05,EIBLET05,COEIET08,KOMCWE07,ROWEET10,ZEZHET11,SDSSDR7}). 
However, as the SDSS Main L6 sample largely overlaps with
the LRG sample, we consider only the first five galaxy samples
among the SDSS Main samples.
Furthermore, while there exist faint and bright LRG sub-samples 
\cite{EIBLET05,ZEEIET05}, we 
combine both LRG samples into a single LRG sample. Finally, the maxBCG samples
representing the most massive clusters are considered with three subsamples,
each of which differs in richness threshold and hence in mass. Our hypothetical
galaxy samples are constructed to represent the observed SDSS galaxy samples
and cover a wide range of mass. The details of the SDSS galaxy samples
are described in Table~\ref{tab:sdss}.

We approximate these SDSS luminosity-bin samples as isolated central galaxies
occupying individual dark matter halos. This approximation is valid for bright 
galaxy samples, as the satellite fraction in typical LRGs is shown to be around
$3-5\%$ \cite{MSCBHB06,ZHZEET09},
while the approximation breaks down for faint galaxy samples,
where a sizable fraction are satellite galaxies that belong to a group or
a cluster of galaxies. However, instead of modeling those galaxy samples
with more free parameters and assumptions, we rely on various methods that
remove satellite galaxies and identify central galaxies (e.g., see 
\cite{RESP09}).
In this way, we can eliminate the uncertainties associated with nonlinear
modeling of galaxies on small scales and focus on the cosmological constraining
power that each galaxy sample represents.

Figure~\ref{fig:samples} describes our model for the SDSS galaxy samples.
The top panel shows the relation between the halo masses and the central
galaxy luminosity. The luminosity of the SDSS galaxy samples are obtained
by using the $K$-corrected rest-frame $r$-band magnitude \cite{BLLIET03},
and their mass ranges $(\Mmin,\Mmax)$ shown as gray boxes are obtained
by matching the observed number density $n_g$ and the 
mean mass $\MB$ (thick vertical lines) from the galaxy-galaxy
lensing measurements \cite{MSKHB06,MASEHI08}. The dashed curve shows the
best-fit relation for the halo mass and the central galaxy luminosity
\cite{ZEZHET11}, obtained by analyzing the observed SDSS galaxy 
clustering within the luminosity range shown as the dotted lines.
We extrapolate the relation (solid) beyond its validity regime to represent
the LRG and the maxBCG samples, but we modify the relation to account for the
fact that the central galaxies at these clusters are just as bright as
the LRGs, while the combined luminosity of clusters is higher.

The four bottom
panels in Figure~\ref{fig:samples} show the number density $n_g$,
the total number $N_\up{tot}$, the average volume $V_\up{avg}$, and the
mean redshift $\bar z$ for the SDSS galaxy samples with their mean mass
and mass range shown as points and horizontal bars, respectively.
The number density and the total number of galaxies are measured quantities,
from which we infer the average volume and the mean redshift for each sample.

Furthermore, we extend our approximation for the galaxy samples
to all mass range and construct continuous mass-bin samples.
The continuous mass-bin samples are composed of halos with 
the mass-bin interval $\Delta\ln M=1.0$ at each mass. 
Figure~\ref{fig:samples}$a$ plots the the number density (solid)
of the continuous mass-bin samples with their mass range (dotted) as a 
function of the mean mass. The mass-bin interval is chosen to match the
observed SDSS galaxy samples (points). The other physical quantities
$N_\up{tot}$, $V_\up{avg}$, and $\bar z$ are shown as the solid curves in
the other panels.

Low mass faint galaxies are abundant
but probe small volume, while the massive luminous galaxies are rare but
measured at larger distance. Therefore, the total number of galaxies are
bounded at very low and high masses. The continuous mass-bin samples
are good approximations to the observed SDSS galaxy samples, with one exception
for the LRG sample, since the LRG sample is obtained with lower limiting
magnitude than the Main galaxy samples. We will use
the continuous mass-bin samples to investigate the cosmological constraining
power at each mass, but we will compare to the nine SDSS mass-bin samples
(points) to make connections to the observations.

\section{Combining Gravitational Lensing and Galaxy Clustering}
\label{sec:combine}
Large-scale galaxy clustering has been measured with high precision, and
its theoretical interpretation is simple; it constrains the product of the
matter fluctuation amplitude $\rms$ and the bias of the galaxy sample.
In this section, we take this constraint from the large-scale clustering 
measurements of each galaxy sample as a base in our cosmological parameter
analysis and combine various gravitational lensing measurements 
to derive further improvements on cosmological parameter constraints in 
a model-independent way. The abundance information is also considered in
conjunction with its mass measurements from gravitational lensing.

To facilitate our understanding of these complementary approaches,
we split the lensing measurements into two regimes (small-scale and
large-scale) and call these two different ways method~I and method~II,
respectively. We investigate in Secs.~\ref{ssec:metI} and~\ref{ssec:metII}
what information each
method can add to the cosmological constraining power. The constraints
from the abundance information (method~III) and from the full analysis 
(method~IV)
of gravitational lensing and galaxy clustering are discussed
in Secs.~\ref{ssec:abun} and~\ref{ssec:full}.

\subsection{Large-Scale Galaxy Clustering}
\label{ssec:lss}
The signal-to-noise ratio of large-scale galaxy clustering
measurements can be determined by the number of independent Fourier modes
obtainable in a survey volume $V_s$ probed by each galaxy sample.
Using the standard mode counting method, we compute the signal-to-noise ratio
of large-scale clustering measurements,
accounting for the shot-noise and the sample variance, as
\beeq
\left({S\over N}\right)^2={1\over2}\int_{\kmin}^{\kmax}{dk\over k} ~
{4\pi k^3V_s\over(2\pi)^3}~\left[{n_gP_0(k)\over 1+n_gP_0(k)}\right]^2~,
\label{eq:snratio}
\eneq
where $\kmin=dk=2\pi/V_s^{1/3}$ and
a factor two accounts for the double counting of Fourier modes
due to the reality of the galaxy fluctuation field. We modeled the
redshift-space galaxy power spectrum on large scales using linear theory 
$P_0(k)=(1+2\beta/3+\beta^2/5)P_g(k)$ and $P_g(k)=b_g^2~\plin(k)$,
where $\beta=f/b_g$ and $f$ is the logarithmic rate of growth.

Given the mass range ($\Mmin$, $\Mmax$) of the SDSS galaxy and the continuous 
mass-bin samples, the galaxy bias factor is 
\beeq
b_g={1\over n_g}\int_{\Mmin}^{\Mmax}dM~{dn\over dM}~b(M)~,
\label{eq:bias}
\eneq
and the mean mass of the samples is
\beeq
\MB={1\over n_g}\int_{\Mmin}^{\Mmax}dM~{dn\over dM}~M~,
\label{eq:mass}
\eneq
where the halo mass function $dn/dM$ and its bias factor $b(M)$ 
are computed by using
the \citet{TIKRET08} and the \citet{TIROET10} fitting formulas, respectively.

\begin{figure}
\centerline{\psfig{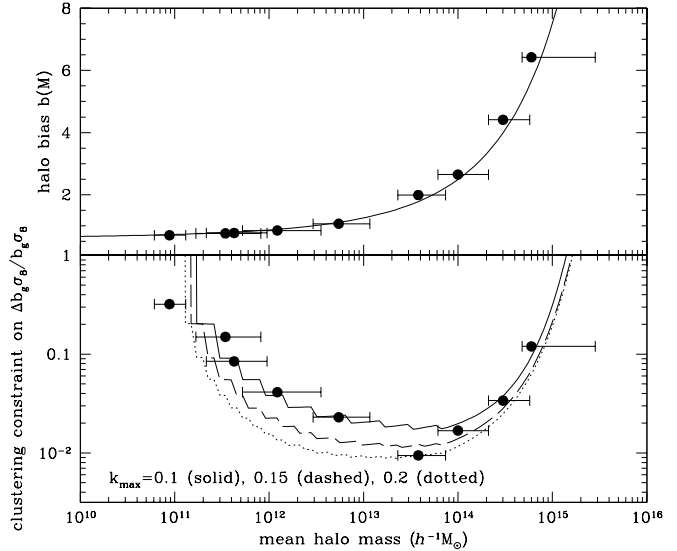}}
\caption{SDSS galaxy bias and clustering constraints. Large-scale galaxy
clustering measurements constrain the combination of the galaxy bias~$b_g$ and
the matter fluctuation amplitude~$\rms$. For the fiducial cosmological model,
two panels show the mean halo bias factor in Eq.~(\ref{eq:bias}) and the 
clustering constraints in Eq.~(\ref{eq:snratio}),
computed at the mean redshift for the SDSS galaxy samples (points)
in Table~\ref{tab:sdss} and at $z=0.1$ for the continuous mass-bin samples
(curves). The clustering constraints are insensitive to the choice of
$\kmin=2\pi/V_s^{1/3}$
but depend on the adopted value of $\kmax$. Due to small volume probed by
less luminous galaxies, the clustering constraints are not smooth at low mass.
}
\label{fig:clustering}
\end{figure}

Figure~\ref{fig:clustering} illustrates the galaxy bias and the clustering
constraints of the galaxy samples. We use Eq.~(\ref{eq:bias}) to model the
galaxy bias factor of each sample. Bias is a monotonic function of mass
and is rather flat in the low mass regime. While the SDSS galaxy samples 
(points) are at different mean redshifts, their bias factors largely agree
with the bias factors of the continuous mass-bin samples at $z=0.1$, as their
redshift range is narrow $z=0.1-0.3$.

The bottom panel shows the clustering constraint on the product $\XX=b_g\rms$ 
(note Eq.~[\ref{eq:snratio}] is the signal-to-noise ratio on square of the
product). Fainter galaxies (low mass halos) have lower shot-noise due to 
larger abundances, though the signal~$\XX$ is rather flat due to the constant 
galaxy bias at low mass. However, the volume probed by these faint 
galaxies is so small as seen in Figure~\ref{fig:samples} that there exist
only few Fourier modes and hence the clustering
 constraint in Figure~\ref{fig:clustering}
is weak at low mass. Especially at low mass, the small volume probed by the
faint
galaxies sets the minimum wavenumber $\kmin$ close to the maximum wavenumber
$\kmax=0.1\hmpci$ we adopted, and
thereby the clustering constraints at low mass are discrete and weak. 

In the high mass regime, where the galaxy samples
correspond to luminous galaxies and clusters, 
the galaxy bias factor continuously increases with mass, and so does the 
clustering signal~$\XX$. However,
while the volume probed by the clusters is 
assumed constant (Fig.~\ref{fig:samples}), the shot-noise of clusters
dramatically increases at high
mass due to the exponential nature of mass function, degrading the clustering
constraints of very massive halos.

Given the volume probed by each galaxy sample and the ratio of their galaxy
power spectrum to the shot-noise, the LRG sample yields the best clustering
constraint. Dashed and dotted curves
in Figure~\ref{fig:clustering} demonstrates
the sensitivity to the maximum wavenumber. For various values of $\kmax$, 
the trend of the clustering constraint in mass 
remains unchanged. While the clustering measurements
are more precise in the nonlinear regime, its theoretical interpretation
becomes more prone to systematic errors from nonlinear evolution.

\subsection{Method I: Small-Scale Galaxy-Galaxy Lensing and 
Large-Scale Clustering}
\label{ssec:metI}

\begin{figure*}
\centerline{\psfig{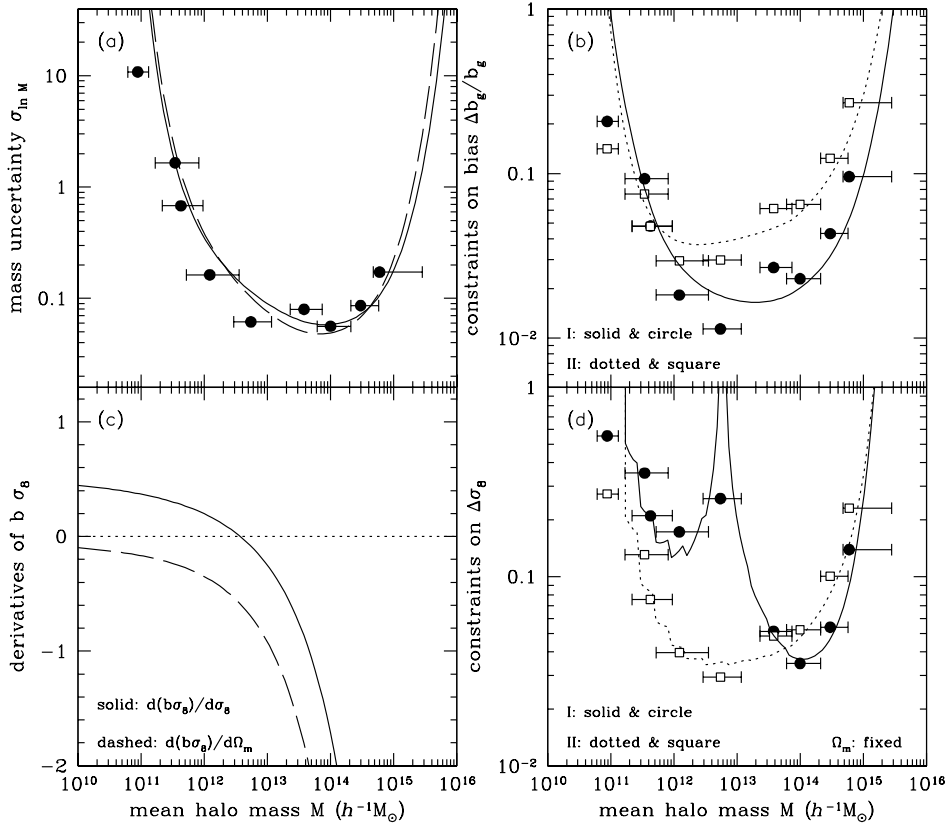}}
\caption{Galaxy-galaxy lensing and large-scale clustering
constraints. ($a$)~Small-scale lensing
measurements constrain the mean mass of the continuous mass-bin (solid)
and the SDSS galaxy (points) samples. The mass 
uncertainties are derived by using the lensing measurements at projected 
separation $R_p=0.2\sim2~\hmpc$, adjusted for each sample 
in proportion to $(M/M_0)^{1/3}$, where $M_0$ is the mean 
mass of the SDSS L4 sample. The dashed line shows the mass uncertainties 
without the mass-dependent adjustment in~$R_p$.
($b$)~Lensing constraints on galaxy bias, using the method~I (solid \& circles)
and the method~II (dotted \& squares).
The mass uncertainties are converted
into uncertainties in galaxy bias (solid)
by using the theoretical prediction of galaxy bias $b(M)$.
Large-scale lensing measurements (dotted) directly constrain the galaxy bias
factor of each sample, and the measurements constraints are derived by using
the signals at $R_p=3\sim60\hmpc$.
Remaining cosmological parameters are held fixed in both cases,
including the mean matter density. 
($c$)~Cosmological parameter sensitivity of clustering constraint $\XX$
in each mass.
($d$)~Constraints on the matter fluctuation amplitude from galaxy-galaxy
lensing and large-scale galaxy clustering. }
\label{fig:lensing}
\end{figure*}

\begin{figure}
\centerline{\psfig{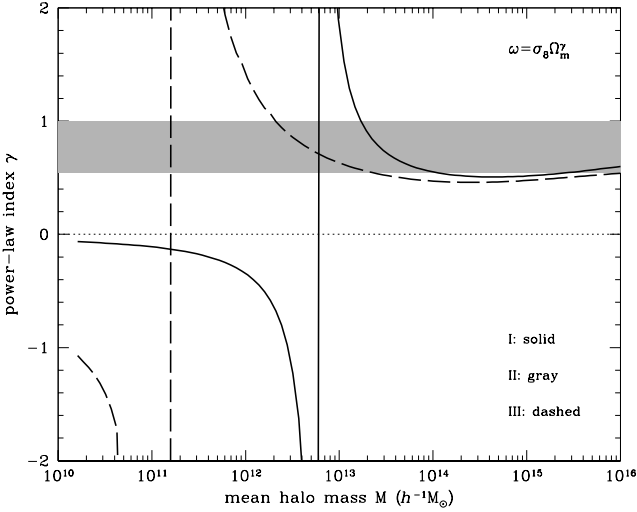}}
\caption{Indices of the cosmological parameter constraints 
from three different methods of combining gravitational lensing and large-scale
galaxy clustering (I: small-scale lensing, II: large-scale lensing,
III: abundance).
The clustering constraints in Fig.~\ref{fig:clustering} are
included in all three methods. Parameter
indices are derived by using Eqs.~(\ref{eq:slopeI}) and~(\ref{eq:slopeIII}).
The value of~$\gamma$ in the method~II depends on the smallest scale
included in the analysis, and its range is plotted as a gray area.}
\label{fig:slope}
\end{figure}

Galaxy-galaxy lensing measurements on small scales provide a robust 
and a model-independent way to estimate the mean mass of the galaxy samples. 
While the mean matter density~$\OM$ may not be well constrained by
gravitational lensing alone, the mean
mass is well constrained, as it is proportional to $\omega_m=\OM h^2$.
So we model two observable constraints, the large-scale galaxy clustering
and the small-scale galaxy-galaxy lensing, by using three independent
physical parameters: 
two cosmological parameters and the mean mass, $\pvec=(\OM,~\rms,~\MB)$. 

The uncertainty in the mass measurements can be computed by adopting the
Fisher information technique. Given the fiducial cosmological parameters, 
we first compute the total number $N_\up{tot}$ of foreground lens galaxies 
shown in Figure~\ref{fig:samples} that will be stacked for galaxy-galaxy
lensing measurements. As a function of angular separation $\theta$ 
from the lens galaxies, the total number of background source galaxies 
for shape measurements is 
\beeq
N_{bg}(\theta)=N_\up{tot}~\nbg~2\pi\theta\Delta\theta~,
\eneq
where the number density of background source galaxies is 
$\bar \nbg=1.2~$arcmin$^{-2}$ in the SDSS and
we use only those galaxies behind the lens galaxies $z_s\geq z_l+0.1$, i.e.,
\beeq
\nbg=\bar\nbg\int_{z_l+0.1}^\infty dz~{d\nbg\over dz}~.
\eneq
The redshift distribution of background source galaxies is normalized, and
we model it with three free parameters ($a,b,c$) as
\beeq
{d\nbg\over dz}={b~c^{1+a\over b}
\over z_0\Gamma\left({1+a\over b}\right)}
\left({z\over z_0}\right)^a\exp\left[-c(z/z_0)^b\right]~,
\label{eq:nbg}
\eneq
where $\Gamma(x)$ is the Gamma function. The mean redshift is
\beeq
\bar z={z_0\over c^{1/b}}{\Gamma\left({2+a\over b}\right)\over
\Gamma\left({1+a\over b}\right)}~,
\eneq
and the median redshift can be obtained from
\beeq
\gamma\left[{1+a\over b},c\left({z_\up{med}\over z_0}\right)^b\right]
={1\over2}\Gamma\left({1+a\over b}\right)~,
\eneq
by solving the incomplete Gamma function~$\gamma(a,x)$.
For the galaxy-galaxy lensing measurements in the SDSS, we adopt
($a,b,c$)=(1.34, 2, 1/2) with the mean redshift $\bar z=1.38z_0=0.42$ and
the median redshift $z_\up{med}=1.31z_0=0.40$ 
from \cite{NAMAET12}.

The total number of background source
galaxies determines the uncertainties in shear measurements
\beeq
\sigma^2_\gamma(\theta)={\sigma^2_\gamma\over N_{bg}(\theta)}
={1\over N_{bg}(\theta)}
\left({\sigma_\up{int}\over\mathcal{R}}\right)^2~,
\eneq
where the intrinsic shape noise is $\sigma_\up{int}=0.37$ and the responsivity
is $\mathcal{R}=1.7$ \cite{MSKHB06}. 
Note that these numbers are observables and
held fixed for cosmological parameter variations.
Therefore, the uncertainties in the mass measurements are
\beeq
{1\over\sigma^2_{\MB}}
=\sum_\theta{1\over\sigma^2_\gamma(\theta)}\left[{\partial\bar\gamma(\theta)
\over\partial\MB}\right]^2~,
\eneq
and the mean shear signal is computed by using the excess surface density 
$\Delta\Sigma$ of the foreground lens galaxy sample as
\beeq
\bar\gamma(\theta)=\int_{z_l+0.1}^\infty dz_s~{d\nbg\over dz_s}~
{\Delta\Sigma(\theta,\MB)\over\Sigma_c(z_l,z_s)}~.
\label{eq:mshear}
\eneq
where the (comoving) critical surface density is
\beeq
\Sigma_c={c^2\over4\pi G}{r_s\over r_lr_{ls}}{1\over1+z_l}~,
\label{eq:CS}
\eneq
and $(r_l,r_s,r_{ls})$ are the comoving angular diameter distances
to the lens, the source, and between the lens and the sources, respectively.
The factor $(1+z_l)$ arises due to our use of comoving angular
diameter distances. The excess surface density $\DS(\theta,M)$ 
on small scales is computed by using the projected NFW density
profile $\Sigma(R)$ \cite{NFW97,WRBR00} as the difference 
$\DS=\bar\Sigma(<\!\!R)-\Sigma(R)$ between the mean
surface density $\bar\Sigma$ interior to the disk and the surface density 
at the same projected radius $R=\theta r_l$.

Figure~\ref{fig:lensing}$a$ shows the uncertainty $\sigma_{\ln\MB}$ in the
mass measurements of the galaxy samples. We limit the small-scale lensing
measurements to the angle, corresponding to the projected separation
$R_p=0.2\sim2\hmpc$ to avoid complications due to the baryonic effects at the
center. The angular range is set for the SDSS L4 sample and is adjusted
for each sample in proportion to its virial radius $\propto M^{1/3}$.
However, the impact of this adjustment is small (dashed). The uncertainty
in the mass measurements is dominated by the total number $N_\up{tot}$
of lens galaxies and hence the total number $N_{bg}$ 
of background source galaxies. The LRG sample exhibits a weaker
constraint $\sigma_{\ln\MB}$ than the L5 or the maxBCG samples, slightly
deviating from the trend shown in the continuous mass-bin samples (solid),
because the mean redshift of the LRG is higher and there exist fewer background
source galaxies for shape measurements.

Now, in order to combine the mass measurements $\MB$ with the 
large-scale clustering constraints $\XX$ and constrain the
cosmological parameters, we perform an error propagation analysis:
\beeq
d\XX={\partial\XX\over\partial\OM}~d\OM+{\partial\XX\over\partial\rms}~d\rms
+{\partial\XX\over\partial\MB}~d\MB~.
\label{eq:dX}
\eneq
With the matter density fixed ($\Delta\OM=0$), the constraint on~$\rms$ 
from the large-scale clustering and the small-scale galaxy-galaxy lensing
measurements is
\beeq
\Delta\rms^2=\left({\partial\XX\over\partial\rms}\right)^{-2}
\left[\sigma_{\XX}^2+\left({\partial\XX\over\partial\MB}\right)^2
\sigma_{\MB}^2\right]~,
\eneq
and this shows how sensitive the large-scale clustering~$\XX$ is to the change
in the mean mass of the galaxy sample and the matter fluctuation amplitude.

The sensitivity to halo mass can be read off from 
Figure~\ref{fig:clustering} as  
$\partial\XX/\partial M=\rms(\partial b_g/\partial M)$~.
Since the galaxy bias factor is  a monotonic function of mass and it is nearly
constant at low mass end, large uncertainty in mass for faint galaxy samples
can only contribute little to the uncertainty in the galaxy bias factor. 
Figure~\ref{fig:lensing}$b$ shows the uncertainty (solid) in bias prediction
$b_g$ constrained by using the small-scale lensing measurements of the mean 
mass. An order of magnitude uncertainty in mass for the SDSS L1$-$L3 samples
translates into fairly good estimates of bias factors, while the galaxy
bias factors are still better constrained for the luminous galaxy samples.

However, an accurate estimate of galaxy bias may be irrelevant in constraining
the cosmological parameter if the product, the clustering constraint~$\XX$,
is independent of the change in the matter fluctuation amplitude.
Figure~\ref{fig:lensing}$c$ plots the sensitivity to the fluctuation 
amplitude (solid) and to the mean matter density (dashed). A high peak of 
a density field becomes less biased as the rms fluctuation amplitude increases.
Therefore, the sensitivity becomes negative at high mass, and there exists
a zero-crossing in the derivative $\partial\XX/\partial\rms$, where the
change in~$\rms$ is compensated by the change in~$b_g$, leaving the clustering
amplitude unchanged. At this mass the method~I cannot give any cosmological 
constraints on the matter fluctuation amplitude $\sigma_8$. 

Figure~\ref{fig:lensing}$d$ illustrates the constraint (solid) 
on~$\rms$ from combining the large-scale clustering and the small-scale
galaxy-galaxy lensing. The insensitivity of the clustering amplitude to the
matter fluctuation amplitude is reflected as no constraint around 
$M\sim5\times10^{12}\hmsun$. In \cite{SEMAET05} 
this method was applied to L4 and L5 
galaxies, which we see are close to this zero crossing where 
no information can 
be extracted. As a result, the derived constraints in \cite{SEMAET05} 
are relatively weak. 
The matter fluctuation amplitude can be best 
constrained by using the LRG or the maxBCG samples around 
$M\simeq10^{14}\hmsun$. At high mass, the low number density $n_g$
(and hence the total number $N_\up{tot}$) is the dominant source of errors 
in measurements of the mean
mass and the clustering amplitude of the galaxy samples.
Since the theoretical prediction of the galaxy bias factor is based on
lensing mass estimates and the galaxy-galaxy lensing measurements are obtained
by averaging over all galaxies with different large-scale environments,
the bias prediction is not subject to the sample variance errors and
is little affected by the systematic errors due to the halo assembly bias 
(e.g., \cite{GSW05,WEZEET06,CRGAWH07}). As long as nearly
all halos at a given mass are included in the sample,
the impact of the halo assembly bias is small.

With two observables~$\XX$ and~$\MB$, we can only constrain two parameters.
To obtain constraints on the cosmological parameter combination, we define 
$\ZZ=\rms^\alpha\OM^\gamma$, and the propagation equation~(\ref{eq:dX}) is
\beeq
d\XX=\left({\partial\XX\over\partial\OM}\right){\OM\over\gamma}{d\ZZ\over\ZZ}
+{\partial\XX\over\partial\MB}~d\MB~,
\eneq
where
\beeq
{d\ZZ\over\ZZ}={\gamma\over\OM}
\left(d\OM+{\alpha\over\gamma}~{\OM\over\rms}~d\rms\right)~
\eneq
and the power-law indices satisfy
\beeq
{\gamma\over\alpha}={\OM\over\rms}{\partial\XX\over\partial\OM}
\left({\partial\XX\over\partial\rms}\right)^{-1}~.
\label{eq:slopeI}
\eneq
Without loss of generality, we let $\alpha\equiv1$, i.e.,
$\ZZ=\rms\OM^\gamma$. 

Figure~\ref{fig:slope} shows the power-law index~$\gamma$ (solid) for the 
method~I of combining the large-scale clustering and the small-scale lensing
measurements. At low mass end, the constraint is insensitive to the mean
matter density $\gamma\simeq0$, as the clustering amplitude is independent
of~$\OM$ (shown as the dashed line in Fig.~\ref{fig:lensing}$c$). The index
becomes infinity, reflecting that there is no constraint on the matter
fluctuation amplitude around the zero-crossing point. Finally, the combination
of large-scale galaxy clustering and small-scale lensing yields constraints
on $\ZZ=\rms\OM^{0.6}$ at high mass.
The fractional uncertainty $\Delta{\ZZ}/\ZZ$ is 
equivalent to the fractional uncertainty $\Delta\rms/\rms$ 
(Fig.~\ref{fig:lensing}$d$) if the mean matter density is known, otherwise one 
can think of this method of constraining $\ZZ$.

\subsection{Method II: Large-Scale Galaxy-Galaxy Lensing and 
Large-Scale Clustering}
\label{ssec:metII}

Galaxy-galaxy lensing on large scales provides measurements of the 
galaxy-matter cross-correlation, and as shown in \cite{YTWZKD06}
and by combining 
it with galaxy clustering one constrains the parameter combination
$\ZZ=\rms\OM^{\gamma}$ with the power-law index $\gamma$ as a function
of scales. On large scales $\gamma=1$, while if one extends the method 
to nonlinear scales 
one finds $\gamma\sim 0.55$ down to $3-4 \hmpc$ \cite{MASLET12}. 
The power-law index of the parameter combination $\ZZ=\rms\OM^\gamma$
is thus between 0.55 and~1, depending on the smallest scale 
included in the analysis, as shown as the gray area in Figure~\ref{fig:slope}.
Here we simply use the linear theory for the method~II, i.e., $\gamma=1$,
in which no modeling of the galaxy bias factor is involved
(hence the method~II is not subject to the halo assembly bias).

Similarly to the 
small-scale lensing constraint on mass, we perform a Fisher matrix calculation
to obtain constraints $\sigma_{\YY}$, where 
$\YY=b_g\OM\rms^2$ is the dependence of the amplitude of galaxy-galaxy 
lensing on 
cosmological parameters. In this case 
the mean shear signal $\bar\gamma(\theta)$
in Eq.~(\ref{eq:mshear}) is obtained by computing the excess surface density 
$\Delta\Sigma$ on large scales with linear theory 
cross-correlation $\xi_{gm}=b_g~\xi_m$.
The lensing signal is summed over the projected separation
$R_p=3\sim60\hmpc$ for all galaxy samples.

The dotted curve in Figure~\ref{fig:lensing}$b$ shows the constraints on
the galaxy bias factor $b_g$ when the cosmological parameters are held fixed
(or the constraints on the large-scale galaxy-galaxy lensing amplitude~$\YY$). 
The method~II works well over a broad range of mass, and the constraint 
from the LRG sample is $\sim8\%$, consistent with \cite{MASLET12}.
However, if we limit the lensing measurements to $R_p=3\sim30\hmpc$ as
the case for the SDSS Main samples, the statistical power would be reduced 
by a factor two, and the LRG sample is best suited for the method~II.

The large-scale clustering constraints are combined 
with the large-scale galaxy-galaxy lensing constraints 
by propagating errors as
\beeq
\left(\begin{array}{c}d\YY\\ d\XX\end{array}\right)=
\left(\begin{array}{ccc}
{\partial\YY\over\partial\OM} & {\partial\YY\over\partial\rms} &
{\partial\YY\over\partial\MB} \\
{\partial\XX\over\partial\OM} & {\partial\XX\over\partial\rms} &
{\partial\XX\over\partial\MB} 
\end{array}\right)
\left(\begin{array}{c}d\OM \\ d\rms \\ d\MB
\end{array}\right)~.
\eneq
With the matter density fixed ($\Delta\OM=0$), the constraint on~$\rms$
from the large-scale clustering and the large-scale galaxy-galaxy lensing
measurements is
\beeq
\hspace{-10pt}
\Delta\rms^2=\left({\partial\XX\over\partial M}{\partial\YY\over\partial\rms}-
{\partial\YY\over\partial M}{\partial\XX\over\partial\rms}\right)^{-2}
\left[\left({\partial\YY\over\partial M}\right)^2\!\!\!\sigma_{\XX}^2
+\left({\partial\XX\over\partial M}\right)^2\!\!\!\sigma_{\YY}^2\right]~.
\eneq
Figure~\ref{fig:lensing}$d$ presents the constraint (dotted) on~$\rms$
from combining the large-scale clustering and lensing measurements.
The best constraints on the matter fluctuation amplitude can be obtained 
from a fairly broad range in mass, covering the SDSS L4 to the maxBCG samples.
The method~II constraints are comparable to the method~I, but they differ
in the mass scale that provides best constraints.

We note that our 
analysis includes only statistical errors and not systematics
(see, e.g., 
\cite{HIMAET04,MHSGET05,HUTAET06,BLMCSE11,NAMAET12}).
For example, currently the dominating source of error is photo-$z$ 
calibration and 
conversion from observed ellipticity to shear, which contribute 
a combined 5\% error on the 
amplitude of galaxy-galaxy lensing signal in the latest 
SDSS analysis \cite{MASLET12}. 
This would suggest that the amplitude cannot be measured to better than
$5\%$ accuracy with the current analysis pipeline (already comparable to
the statistical errors in the SDSS), and hence the constraint
becomes inflated to $\Delta\rms\sim0.05$. 
The same systematic error would also apply to the shear-shear analysis, 
which results in $\Delta\rms\sim0.05$ in the linear regime and somewhat
weaker systematic error $\Delta\rms\sim0.03$ in the quasilinear regime,
as it scales as $\rms^3$ \cite{JASE97}.
However, an additional 5\% shear-calibration uncertainty 
in the method~I sets a floor at $\sim7\%$ in mass 
measurements (since $M \sim \Sigma^{3/2}$), which degrades
the $\rms$-constraint to $\Delta\rms\sim0.04$. 
This suggests that method I is less 
sensitive to the overall calibration error than method II 
and shear-shear method. 
This may be an important consideration for the future surveys too.

\begin{figure}
\centerline{\psfig{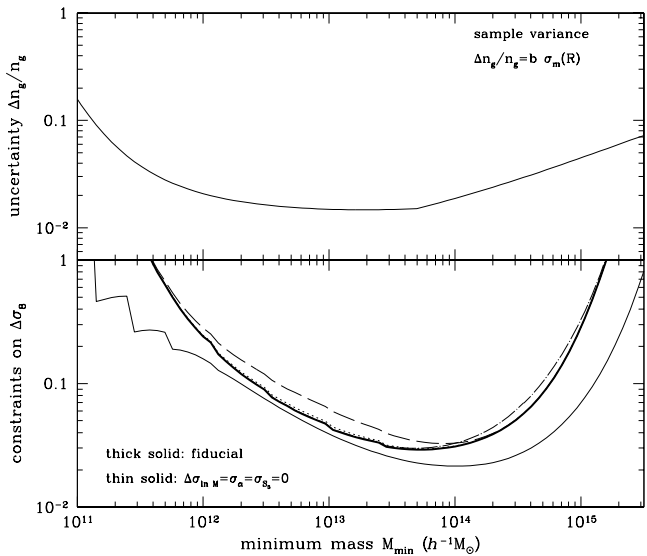}}
\caption{Additional abundance constraints. Upper panel shows the sample 
variance in abundance due to
the limited volume probed by the thresholded samples. Bottom panel shows
the constraints on the matter fluctuation amplitude by adding additional
abundance information.
The mean mass and the clustering constraints are combined to constrain the
minimum mass $\Mmin$, the log-normal scatter $\sigM$, and the matter 
fluctuation amplitude $\rms$ (thick solid). We assume $\sigM=0.5$ in the
fiducial model.
In addition, we consider two systematic errors: invisible halos
and skewness ($S_3\neq0$) in the mass-observable relation.
If some fraction ($\alpha_g>0$) of halos are
devoid of galaxies, they may drop out of the sample, acting as a systematic
error.
Dotted ($\sigma_{\alpha_g}=0.1$), dashed ($\sigma_{\alpha_g}=0.3$), and
dot-dashed ($\sigma_{S_3}=5.0$) curves show the degradation 
in $\Delta\rms$ due to systematic errors, 
if we marginalize over those parameters in our modeling. 
The constraint can be greatly improved (thin solid),
if the log-normal scatter, the invisible fraction, and the skewness
are perfectly known ($\Delta\sigM=\Delta\alpha_g=\Delta S_3=0$).}
\label{fig:abundance}
\end{figure}

\subsection{Method III: Additional Abundance Information}
\label{ssec:abun}
Now we add the abundance information in the analysis and investigate its
cosmological constraining power. Since the number density $n_g$ of 
galaxy samples is rather sensitive to the mass-observable relation, we
consider in this case thresholded galaxy samples, instead of luminosity-bin 
samples. As our fiducial model, we assume that the thresholded galaxy samples 
are described by two parameters, the minimum threshold mass $\Mmin$ 
and the log-normal scatter $\sigM$ of the mass-observable relation,
given the cosmological parameters: $\pvec=(\OM,\rms,\Mmin,\sigM)$.
We set $\sigM=0.5$ in the fiducial model.

As our observables of the thresholded galaxy samples,
we consider the abundance $\NN$ of the galaxy sample,
the mean mass $\MB$ of the galaxy sample from small-scale lensing 
measurements, ßand the large-scale clustering $\XX$: $\bdv{\OO}=(\NN,\MB,\XX)$.
The number density of the thresholded galaxy samples can be computed, 
accounting for the scatter in the mass-observable relation, as
\beeq
\NN=\int_0^\infty dM~{dn\over dM}~{1\over2}~{\erfc}
\left[{\ln\Mmin-\ln M\over\sqrt{2}\sigM}\right] ~,
\label{eq:thrsn}
\eneq
where erfc($x$) is the complementary error function. The mean mass and the 
bias of the thresholded samples are then
\bear
\label{eq:thm}
\MB&=&{1\over \NN}\int_0^\infty dM~M~{dn\over dM}
~{1\over2}~{\erfc}\left[{\ln\Mmin-\ln M\over\sqrt{2}\sigM}\right] ~, \\
b_g&=&{1\over \NN}\int_0^\infty dM~b(M)~{dn\over dM}
~{1\over2}~{\erfc}\left[{\ln\Mmin-\ln M\over\sqrt{2}\sigM}\right] ~. \nonumber
\enar

We use the same mass-observable relation for the central galaxies of the
thresholded samples as described in Figure~\ref{fig:samples}, in computing
the volume probed by the thresholded galaxy samples. Measurement uncertainties
in the mean mass and the large-scale clustering amplitude are therefore
obtained in the same way. While the measurement uncertainty in abundance is 
practically zero in observation, our theoretical prediction~$\NN$ in
Eq.~(\ref{eq:thrsn}) is based on the globally averaged mass function, which 
differs from the measurements due to the sample variance 
(see, e.g., \cite{HUKR03}). Therefore, we treat the sample variance in 
abundance as the measurement uncertainty $\sigma_{\NN}$ in abundance,
and for a galaxy sample covering the volume~$V$ the sample variance is 
computed as
\beeq
\left({\Delta\NN\over\NN}\right)^2={\AVE{\NN^2}-\bar\NN^2\over\bar\NN^2}
=b_g^2\sigma^2_m(R)~,
\eneq
where $R=(3V/4\pi)^{1/3}$ and $\sigma_m(R)$ is the rms matter fluctuation
smoothed by a top-hat filter with~$R$.

The upper panel in Figure~\ref{fig:abundance} shows the fluctuation in the
abundance~$\Delta\NN/\NN$ due to the limited volume probed by the thresholded
galaxy samples. As the volume coverage increases with the minimum threshold
mass (or brighter galaxies), the sample variance decreases accordingly, and the
observed abundance approaches to the global mean value to a percent level
around $\Mmin\sim10^{14}\hmsun$. The upturn at $M>10^{14}\hmsun$ reflects
the fact that the central galaxies of more massive clusters are not brighter, 
while they are more biased.

Using the matrix form between the observables~$\bdv{\OO}$ and the 
parameters~$\pvec$, a general error propagation analysis can be performed
as
\beeq
d\bdv{\OO}={\partial\bdv{\OO}\over\partial\pvec}~d\pvec~,
\eneq
and assuming $\Delta\OM=0$ the constraint on~$\rms$ from the thresholded
galaxy samples can be obtained as
\bear
\label{eq:abund}
\Delta\rms^2&=&{1\over D^2}\Bigg[
\left({\partial\XX\over\partial\sigM}{\partial\MB\over\partial\Mmin}-
{\partial\XX\over\partial\Mmin}{\partial\MB\over\partial\sigM}\right)^2
\sigma^2_{\NN} \nonumber \\
&&+\left({\partial\XX\over\partial\sigM}{\partial\NN\over\partial\Mmin}-
{\partial\XX\over\partial\Mmin}{\partial\NN\over\partial\sigM}\right)^2
\sigma^2_{\MB}\nonumber \\
&&+\left({\partial\MB\over\partial\sigM}{\partial\NN\over\partial\Mmin}-
{\partial\MB\over\partial\Mmin}{\partial\NN\over\partial\sigM}\right)^2
\sigma^2_{\XX}\Bigg]~,\nonumber
\enar
where the determinant of the matrix with $\pvec=(\rms,\Mmin,\sigM)$ is
\bear
D&=&\left|{\partial\bdv{\OO}\over\partial\pvec}\right| 
=\left({\partial\NN\over\partial\rms}\right)
\left({\partial\XX\over\partial\sigM}{\partial\MB\over\partial\Mmin}
-{\partial\XX\over\partial\Mmin}{\partial\MB\over\partial\sigM}\right) 
\nonumber\\
&&-\left({\partial\MB\over\partial\rms}\right)
\left({\partial\XX\over\partial\sigM}{\partial\NN\over\partial\Mmin}
-{\partial\XX\over\partial\Mmin}{\partial\NN\over\partial\sigM}\right) 
\nonumber\\
&&+\left({\partial\XX\over\partial\rms}\right)
\left({\partial\MB\over\partial\sigM}{\partial\NN\over\partial\Mmin}
-{\partial\MB\over\partial\Mmin}{\partial\NN\over\partial\sigM}\right)~.
\nonumber
\enar
The bottom panel in Figure~\ref{fig:abundance} shows the constraint 
(thick solid) on~$\rms$ by adding the abundance information. 
Due to the exponential
sensitivity, the derivatives of the abundance $\partial\ln\NN/\partial\ln\pvec$
are larger than the derivatives of the clustering amplitude and the mean mass,
and the sample variance error in the
abundance is always an order-of-magnitude smaller than the other uncertainties.
Consequently, the uncertainties
in the clustering amplitude and the mean mass dominate the error budget in
$\Delta\rms$ at all mass range, and
the uncertainty in the matter fluctuation amplitude is set by
the uncertainty in the clustering amplitude at high mass and the uncertainty
in the mean mass at low mass.

Significant improvements (thin solid) can be made in constraining the 
matter fluctuation amplitude using the abundance information, if we have
extra information about the log-normal scatter in the mass-observable
relation. At a mass of $10^{14}\hmsun$ this leads to a factor of two 
improvement 
in the error. This shows the importance of knowing the scatter in the 
cluster abundance method. However, obtaining this scatter from other
observations is difficult, since other observables have different
distributions and associated scatters. Our 
approach of extracting it from the clustering 
is more conservative, as we discuss in more detail below. 
Though the constraints are still dominated by the uncertainties in 
the clustering amplitude and the mean mass, the sample variance becomes
non-negligible in this limit.
The power-law index of the parameter combination $\ZZ=\rms\OM^\gamma$ can
be obtained by using the matrix form as
\beeq
\gamma={\OM\over\rms}{D\left[\pvec=(\OM,\Mmin,\sigM)\right]\over 
D\left[\pvec=(\rms,\Mmin,\sigM)\right]}~,
\label{eq:slopeIII}
\eneq
and Figure~\ref{fig:slope} shows the power-law index~$\gamma$ (dashed)
for the method~III. A unique signature is again there exist a mass range,
where no constraint on the matter fluctuation amplitude can be derived
($\gamma\simeq\infty$).
Since the change in the matter fluctuation amplitude
results in not only the increase in the overall abundance of halos but also
the distortion in the halos mass function, there exists a mass range
$M\sim10^{11}\hmsun$,
at which no constraint on $\rms$ can be obtained from the abundance.
At high mass, the thresholded samples approach the mass-bin samples, and the
constraints becomes $\omega\simeq\rms\OM^{0.6}$ as in the method~I.

\begin{figure*}
\centerline{\psfig{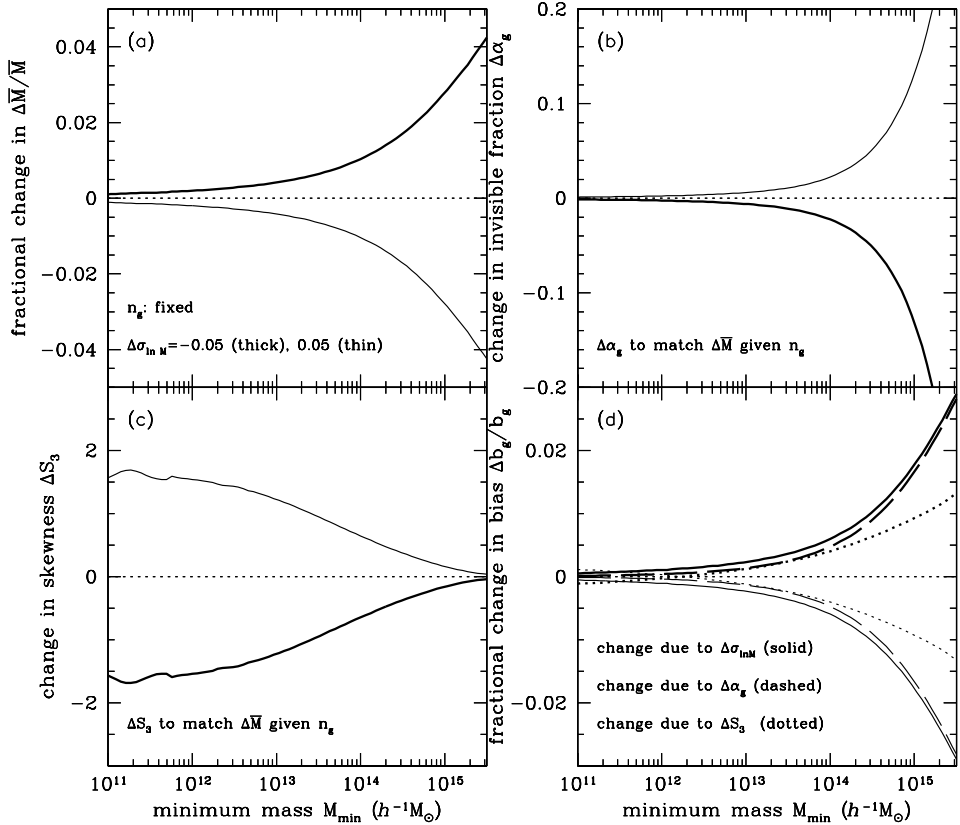}}
\caption{Degeneracy in additional parameters of the thresholded samples.
$(a)$~The fractional change in the mean mass $\MB$ due to the change in 
the log-normal scatter $\sigM$. 
The minimum threshold mass $\Mmin$ is adjusted to compensate for the 
resulting change in abundance in all panels, i.e., $\NN$ is fixed as the
value in the fiducial model.
$(b)$~The change in the invisible halo fraction~$\alpha_g$
required to match $\Delta\MB$ in Panel~($a$). While negative values of
$\alpha_g$ are plotted for illustration purposes, only non-negative values
of $\alpha_g$ have physical meaning. $(c)$~The change in the 
skewness~$S_3$ required to match $\Delta\MB$ in Panel~($a$). The abundance
is again fixed as in the fiducial model ($S_3=0$).
$(d)$~The fractional change in the galaxy bias factor due to the change
$\Delta\sigM$, $\Delta\alpha_g$, and $\Delta S_3$ in each panel.
The invisible fraction parameter~$\alpha_g$ is largely degenerate with the
log-normal scatter $\sigM$, while the skewness~$S_3$ is only degenerate
below $\Mmin<10^{14}\hmsun$. }
\label{fig:syserr}
\end{figure*}

\subsubsection{Systematic Errors: Invisible Halos and Skewness in PDF}
\label{ssec:sys}
Although the abundance information is a powerful tool to probe cosmology due to
its exponential sensitivity at high mass, it is equally sensitive to systematic
errors in theoretical modeling of the abundance. Here we consider two simple
cases for potential systematic errors: 
Invisible halos and skewness in the probability distribution of the 
mass-observable relation. Invisible halos would appear to be a killer for 
this method, since with lensing and abundance information one would never 
be able to tell that some clusters are not present in the sample. However, 
with clustering information one can tell the difference. For example, if we 
assume for simplicity that there is no scatter then in a given model at a given 
abundance we measure all of the most massive halos at that abundance. If a fraction 
of these massive halos are dark, they are not in the catalog and need to be 
replaced with less massive halos to reach the same abundance. But these less massive 
halos also have lower bias, and the two cases can be distinguished if we include 
clustering information in addition to lensing and abundance. In this sense dark halos 
act in the same way as the scatter between mass and observable. Our goal 
in this subsection is to explore 
this in more detail to see if there is any residual difference between the two. 

The first case is that some fraction~$\alpha_g$ of halos at each mass fail
to form galaxies and
simply drop out of the observed galaxy samples. These invisible halos would
act as a systematic error in our cosmological parameter analysis.
In the presence of the invisible fraction, the halo mass
function should be modified as $(1-\alpha_g)(dn/dM)$ in computing the number
density in Eq.~(\ref{eq:thrsn}), while the mean mass and the bias factors
in Eq.~(\ref{eq:thm}) remain unaffected. The effect of the invisible fraction
on the mean mass and the bias factor arises solely from the change in
$\Mmin$, if the number density is held fixed.

The second case of systematic errors is a non-Gaussian
probability distribution in the mass-observable relation. 
The thresholded galaxy samples
are modeled by using the threshold mass~$\Mmin$ and the log-scatter~$\sigM$,
assuming that the scatter between the observed mass and the true mass is
a Gaussian. Now we consider a deviation from the Gaussian assumption.
A non-Gaussian probability distribution function with known cumulants
$\kappa_n$ can be constructed by using a Gaussian distribution as
\cite{MATSU99}
\beeq
P_{nG}(\delta)=\exp\left[\sum_{n=3}^\infty{\kappa_n\over n!}
\left(-{d\over d\delta}\right)^n\right]P_G(\delta,\mu,\kappa_2)~,
\eneq
where the mean~$\mu$ and the variance~$\kappa_2$ of the Gaussian distribution
are set equal to those in the non-Gaussian distribution. 
Using the Edgeworth expansion, we only keep the leading order correction to
the Gaussian distribution,
\beeq
P_{nG}(\delta)\simeq P_\up{G}(\delta,\mu,\kappa_2)\left[1+{\kappa_2\over6}
S_3H_3\left({\delta\over\kappa_2}\right)\right]~,
\label{eq:non}
\eneq
where the third-order 
Hermite polynomial is $H_3(x)=x^3-3x$ and the skewness is 
$S_3=\kappa_3/\kappa_2^2$. In the presence of skewness ($S_3\neq0$), 
the change in the number density is therefore
\bear
\hspace{-5pt}
\Delta\NN&=&{S_3\sigM\over12}\sqrt{2\over\pi}\int_0^\infty dM~{dn\over dM}
\left[{(\ln\Mmin-\ln M)^2\over\sigM^2}-1\right] \nonumber \\
&&\times\exp\left[-\left({\ln\Mmin-\ln M\over\sqrt{2}\sigM}\right)^2\right]~,
\enar
and similar calculations can be performed for
the mean mass and the bias of the thresholded samples.

For intuitive understanding of the degeneracy between these additional 
parameters and the fiducial model parameters, we compute the change in
our observables $\bdv{\OO}=(\NN,\MB,\XX)$, given cosmological parameters
and the most well-measured abundance fixed. Figure~\ref{fig:syserr}$a$ shows
the fractional change in the mean mass~$\MB$ due to the change in
log-normal scatter~$\sigM$. In order to compensate for the change in the
abundance due to the increase  in $\sigM$, the minimum threshold 
mass $\Mmin$ has to increase. However, the dilution of low mass halos
results in decrease in the mean mass (thin solid). The impact is smaller
at low $\Mmin$ as the mass function flattens and the mean mass is dominated
by $M\gg\Mmin$.

Same effects can be achieved by introducing the invisible halo fraction
$\alpha_g\neq0$ or changing the skewness of the probability distribution 
$S_3\neq0$, without changing the log-normal scatter $\sigM$. 
For a fixed abundance, the invisible halo fraction $\alpha_g>0$ 
(thin solid) in 
Figure~\ref{fig:syserr}$b$ needs lower minimum threshold mass and hence 
provides the same effect of reducing the mean mass. Since the magnitude of the
change in~$\alpha_g$ is directly proportional to the change in abundance,
a fixed $\Delta\sigM$ yields progressively small $\Delta\alpha_g$
at low mass.

A positive skewness (thin solid) in Figure~\ref{fig:syserr}$c$ puts more
weight on lower mass and brings more low mass halos into the thresholded 
sample that mimicks the effect of the log-normal scatter on abundance. 
Similar to the log-normal scatter, the dilution of low mass halos decreases
the mean mass. However, since thresholded samples are used, the change
in the probability distribution affects the samples significantly at high
mass, but little at low mass. Therefore, the degree of
change allowed for the skewness can be large $|S_3\sigM|>1$, especially at
low threshold mass.

Figure~\ref{fig:syserr}$d$ illustrates the level of degeneracy of 
these additional parameters. For a fixed abundance, while the change in the
mean mass can be masked by the change $\alpha_g$ or $S_3$, the change in the
galaxy bias factor responds in a different way. The invisible fraction
(dashed) is largely degenerate with the change (solid) in the log-normal
scatter, such that the invisible halo fraction needs to be constrained to
the level seen in Figure~\ref{fig:syserr}$b$. While the skewness (dotted)
in the probability distribution yields somewhat different effects,
the hyper-sensitivity to $S_3$ at high mass puts a very stringent 
requirements on $S_3$.

The impact of these systematic errors on the $\rms$-constraint is presented
in Figure~\ref{fig:abundance}. The fiducial model assumes no invisible
halos ($\alpha_g=0$) or skewness ($S_3=0$). However, with incomplete knowledge
on the invisible halo fraction~$\alpha_g$ 
or the skewness~$S_3$, we marginalize over
each of these parameters, and the constraint is inflated:
$\sigma_{\alpha_g}=0.1$ (dotted), $\sigma_{\alpha_g}=0.3$ (dashed),
$\sigma_{S_3}=5.0$ (dot-dashed).
This demonstrates the level of constraints 
on $\alpha_g$ and $S_3$ required to avoid a significant degradation on 
$\Delta\rms$. As shown in Figure~\ref{fig:syserr}, the impact of the
skewness is larger at high mass, and the impact of the invisible fraction
is larger at low mass, but neither are very important at this level of precision, although 
they may become more important for the future data sets.
Since the method~III is essentially based on the method~I, the same argument
applies to the method~III, regarding the halo assembly bias: provided that
all halos of a given mass are included in the sample, the systematic errors
due to the halo assembly bias should be minimal. Moreover, the
assembly bias effect is small at high mass.

\begin{figure}
\centerline{\psfig{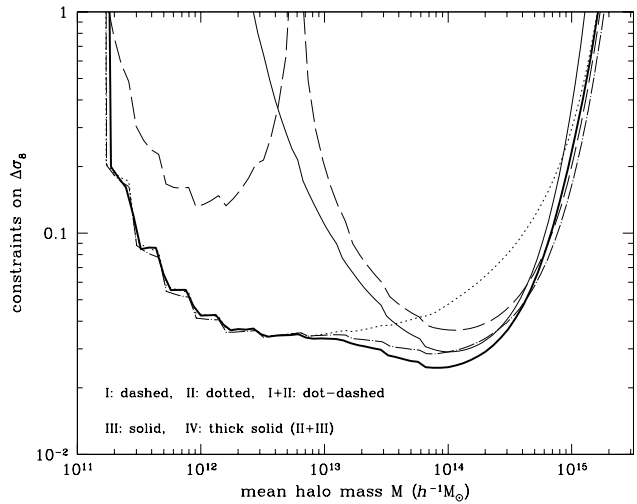}}
\caption{Constraints on the matter fluctuation amplitude. Various curves show
as a function of the mean mass of each sample, the constraints on
$\Delta\rms$ from three different methods (I, II, III). 
The dot-dashed curve shows the combination of method~I and~II, representing
the constraints by combining the gravitational lensing on all scales and the
large-scale clustering. The thick solid curve (method IV) shows the constraints
by adding the abundance information to the gravitational lensing and
the large-scale clustering constraints.}
\label{fig:final}
\end{figure}

\subsection{Combined Analysis of Galaxy-Galaxy Lensing 
and Large-Scale Clustering}
\label{ssec:full}
For easy comparison of various methods for combining galaxy-galaxy lensing
and galaxy clustering,
Figure~\ref{fig:final} summarizes the constraints on the matter fluctuation
amplitude derivable from the SDSS galaxy samples.
The constraint (thin solid) from the method~III appears different 
from that in Figure~\ref{fig:abundance}, as it is plotted in terms of the
mean mass of the galaxy samples, rather than the minimum mass. Since the
method~III makes use of the mass measurements in addition to the abundance
information, it is effectively built upon the method~I. Therefore, the 
constraints (thin solid) of the method~III encompass the constraints (dashed)
of the method~I, although little difference arises due to the difference 
between the thresholded and the mass-bin samples, especially at low mass.

The dot-dashed curve shows the combined constraints of the large-scale
galaxy clustering and the galaxy-galaxy lensing on all scales 
(method~I and~II). The constraints are obtained by adding the constraints
from both methods in inverse quadrature. In this way, the large-scale 
clustering constraints in both methods are doubly counted. However, since
the uncertainty in the clustering measurements is smaller than that in 
the lensing measurements in method~II on large scales (and vice versa on
small scales) and this trend is reversed in method~I,
the double counting of large-scale clustering constraints
affects the final constraints little.

While the method~III (solid) provides the best constraints among the three
different methods, 
the combination of galaxy-galaxy lensing and
large-scale clustering measurements (method I+II: dot-dashed) 
provides equally strong
constraints, while it is less subject to systematic errors that may be present
in the method~III. 
Finally, all three methods can be combined,
 representing the full constraints (thick solid)
by using the galaxy-galaxy lensing and large-scale clustering measurements.
Compared to the method~I+II,
the abundance can bring additional information and tighten the constraints,
albeit not much, around the mass range from the SDSS L4 to the maxBCG samples.

\begin{table*}
\caption{Specifications of the future galaxy surveys adopted for forecasts. 
The sky coverage $\Omega_\up{survey}$ is in units of square degrees,
and the clustering measurements are assumed to be at the mean redshift 
$z_\up{cl}$. The limiting magnitude is $m_\up{lim}$ in $r$-band,
and the intrinsic shear noise is 
$\sigma_\gamma$. The mean number $\bar\nbg$ of background source galaxies 
is per arcminute squared, and 
the redshift distribution parameters $(a,b,c,z_\up{med})$ 
of the background source galaxies are described
in Eq.~(\ref{eq:nbg}). The last column indicates if the surveys are equipped
with spectroscopic redshift measurements. 
}
\begin{tabular}{ccccccccccccccccccccc}
\hline\hline
Survey && $\Omega_\up{survey}$ && $z_\up{cl}$ && $m_\up{lim}$ && $\bar\nbg$ &&
$\sigma_\gamma$ && $a$ && $b$ && $c$ && $z_\up{med}$ && spec-$z$ \\
\hline
SDSS && 8000 && 0.1 && 22.5 && 1.2 && 0.22 && 1.34 && 2 && 1/2 && 0.4 && Y \\
DES  && 5000 && 0.5 && 24.0 && 12.0 && 0.20 && 2.0 && 1.5 && 1 && 0.8 && N \\
DES+BigBOSS && 1000 && 0.5 && 24.0 && 12.0 && 0.20 && 2.0 && 1.5 && 1 && 0.8 && Y \\
Euclid && 20,000 && 0.7 && 24.0 && 40.0 && 0.20 && 2.0 && 1.5 && 1 && 1.0 && Y \\
LSST && 18,000 && 0.9 && 27.0 && 45.0 && 0.20 && 2.0 && 1.5 && 1 && 1.3 && N \\
\hline
\end{tabular}
\label{tab:future}
\end{table*}

\section{Future Galaxy Surveys}
\label{sec:fut}
Here we extend our analysis performed in Sec.~\ref{sec:combine} to future
galaxy surveys like the DES\footnote{http://www.darkenergysurvey.org},
BigBOSS\footnote{http://bigboss.lbl.gov}, 
Euclid\footnote{http://sci.esa.int/euclid}, 
and the LSST\footnote{http://www.lsst.org}, and we
forecast constraints derivable from these future surveys by combining galaxy
clustering and gravitational lensing measurements. Compared to our previous 
investigation of the SDSS, the key difference is that
these galaxy surveys have significantly lower threshold in flux, 
and they can observe fainter galaxies, 
thereby probing larger volume at higher redshift and providing higher
statistical constraining power.

In addition to these simple scaling changes, two quantitative differences
arise in the future galaxy surveys. Photometric surveys like the DES and the
LSST will have a limited capability to map the galaxy positions in three
dimensional space. Provided that there exist a small number of galaxies
with spectroscopic redshift measurements for photometric redshift calibration,
galaxy-galaxy lensing analysis can be performed without much degradation,
compared to the surveys with full spectroscopic capacity.
However, a full three-dimensional analysis of galaxy clustering will be
unavailable in these photometric surveys. Moreover, 
we consider a hypothetical deep-imaging and spectroscopic survey over
1000~deg$^2$, which may be available from the overlapping region
of the DES and the BigBOSS.
This combination may allow for both gravitational
lensing and galaxy clustering analyses. The BOSS and the BigBOSS alone
are not considered here as they are mainly spectroscopic surveys without
imaging and hence there are too few background source galaxies
for gravitational lensing analysis in these surveys.

The other difference in the future surveys
and possibly the major improvement over the SDSS is the availability of the
cosmic shear measurements. A large number of background source galaxies
in the future surveys enable high precision measurements of the
shear-shear auto-correlation, which have been measured only in a very
limited region of deep imaging in the SDSS \cite{HUHIET11,HUEIET11,LIDOET11}.
In this section, we account for these changes by using the galaxy
angular power spectrum in computing the clustering constraints 
for photometric surveys and by
comparing the improved constraints in the future surveys to those derivable
from the cosmic shear measurements. Table~\ref{tab:future} summarizes
the specifications of the future galaxy surveys we adopt in this paper.

\begin{figure*}
\centerline{\psfig{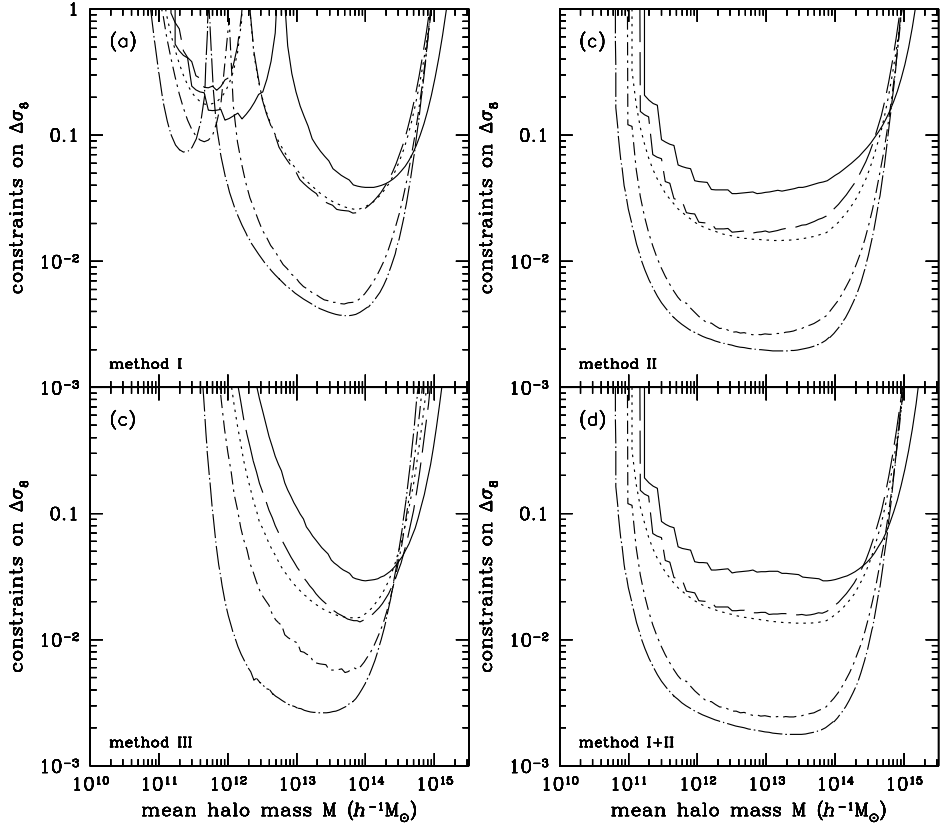}}
\caption{Improvements of combining the large-scale galaxy clustering 
and the galaxy-galaxy
lensing (methods I, II, III, and I+II) 
on the $\rms$-constraints in future galaxy 
surveys. Each galaxy survey is indicated by various curves: SDSS (solid),
DES (dotted), DES+BigBOSS (dashed), Euclid (short dot-dashed), and LSST
(long dot-dashed). Compared to the SDSS (solid), the major 
improvements of the future
galaxy surveys result from the larger survey volume at higher redshift.
The specifications of the future galaxy surveys 
are summarized in Table~\ref{tab:future}.}
\label{fig:future}
\end{figure*}

\subsection{Combining Galaxy-Galaxy Lensing and Large-Scale
Galaxy Clustering}
\label{ssec:combine}
While it is difficult to know which galaxy samples with what properties
will be measured in the future surveys, it matters little to our present
purposes of using them for gravitational lensing and galaxy clustering
measurements. Here we approximate the galaxy samples in the future surveys
as the mass-bin halo samples
and compute the physical quantities plotted in Figure~\ref{fig:samples}.
The number density $n_g$
of the galaxy samples is computed by adopting the same
mass-bin interval $\Delta\ln M=1.0$ as in the SDSS, but the resulting
number density in the future surveys is lower 
at high mass, as the mass function is
computed at higher redshift $z_\up{cl}$ described in Table~\ref{tab:future}.

We use the same SDSS relation between the central galaxy luminosity and its
halo mass as in Figure~\ref{fig:samples}$a$ (solid)
to compute the mean luminosity
distance that these galaxy samples can be measured in the future surveys.
The mean luminosity distances are larger for the galaxy samples in the future
surveys than those with the same absolute luminosities (or the same masses)
in the SDSS, because the limiting flux $f_\up{lim}$
is lower in the future surveys.
The average volume probed by the galaxy samples is therefore 
$V_\up{avg}(z)\propto f^{-3/2}_\up{lim}/(1+z)^3$, where the limiting flux
is related to the limiting magnitude
$m_\up{lim}=-2.5\log_{10} f_\up{lim}+m_0$ with additive constant $m_0$.
We compute the mean luminosity distances and the average volumes for each
sample in the future surveys
by scaling those quantities in the SDSS.
We can then readily obtain
the total number $N_\up{tot}$ and the mean redshift $\bar z$ of
the galaxy samples in the future surveys.\footnote{We note that the limiting
magnitudes of the future surveys in Table~\ref{tab:future} represent those
for the photometric imaging, and those for the spectroscopic measurements
are shallower. However, since they are scaled with the SDSS, we suspect no
substantial difference in our projection. More importantly,
our projection for the
future surveys should be taken with caution, as many uncertain factors
can affect the results presented here.}

For photometric surveys like the DES and the LSST, we compute the 
signal-to-noise ratio of the angular clustering measurement, similarly to 
Eq.~(\ref{eq:snratio}), but
accounting for the lack of three-dimensional information as
\beeq
\left({S\over N}\right)^2=\sum_{l=l_\up{min}}^{l_\up{max}}(2l+1)
{f_\up{sky}\Delta l(C^g_l)^2\over2\left[C^g_l+C^n_l\right]^2}~,
\label{eq:asn}
\eneq
where $f_\up{sky}$ is the fraction of the sky covered by the survey
and the noise power spectrum is $C_l^n=4\pi f_\up{sky}/N_\up{tot}$.
Adopting the Limber approximation and assuming that the galaxy samples are
uniformly distribution in a narrow redshift range $z_\up{cl}\pm\Delta z$,
we compute the galaxy angular power spectrum as
\beeq
C_l^g=\int dz~{H(z)\over r^2}\left({dn_g\over dz}\right)^2
P_g(k)={1\over r^2\Delta r}P_g\left(k={l\over r}\right) ~,
\eneq
where $\Delta r$ is the width of the redshift bin $\Delta z=0.05$. To be 
consistent with our calculations in spectroscopic surveys,
the range of the angular multipoles in Eq.~(\ref{eq:asn})
is obtained by using the Limber relation $k=l/r$, given the range of 
wavenumber ($\kmin$, $\kmax$). In addition, we adopt $\kmax=0.15\hmpci$
to take advantage of the fact that the structure probed by the future surveys 
remains in more linear regime than that probed by the SDSS.

Figure~\ref{fig:future} describes the improvements of the matter fluctuation
constraints over the SDSS
by using the three different methods of combining the large-scale
galaxy clustering and the galaxy-galaxy lensing measurements in various
future surveys (different curves).
The key improvements result from the larger survey volume
and the larger number of background source galaxies at higher redshift,
although the reduced number density increases the shot noise contribution.

\begin{figure*}
\centerline{\psfig{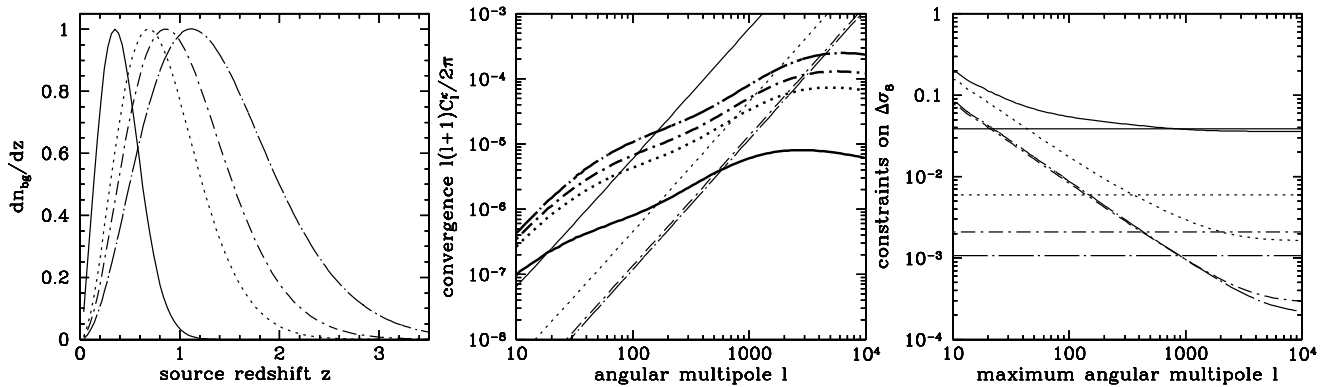}}
\caption{Constraints on the matter fluctuation amplitude from the cosmic
shear measurements in future surveys.
Various curves represent different galaxy surveys
as in Figure~\ref{fig:future} and Table~\ref{tab:future}:
SDSS (solid), DES (dotted), Euclid (short dot-dashed), and LSST (long
dot-dashed).
{\it Left:} The redshift distribution of the
background source galaxies. {\it Middle:} The angular convergence
power spectrum
(thick) and the noise power spectrum (thin). {\it Right:} The constraints
on the matter fluctuation amplitude as a function of the maximum angular
multipole. For comparison only, we included the SDSS (solid) for the
cosmic shear measurements, although it is hard to measure in the SDSS.
Horizontal lines represent the non-Gaussian contributions of the lensing 
trispectrum and the sample variance \cite{COHU01,FOMSWG,SAHAET09}, 
both of which dominate the error budget on small scales.}
\label{fig:shear}
\end{figure*}

Figures~\ref{fig:future}$a$ and~\ref{fig:future}$b$ show the constraints
on $\rms$ by combining the large-scale clustering and the galaxy-galaxy
lensing measurements
on small and large scales, respectively. In addition to the 
improvements in the clustering constraints, a larger number density of
background source galaxies at higher redshift enables higher precision 
lensing measurements in the future surveys. Indeed, the contribution of
the clustering constraints in the method~II are weaker than the contribution
of the large-scale galaxy-galaxy lensing constraints, whereas in the SDSS
the trend is opposite. However, in the method~I the contribution of the
small-scale galaxy-galaxy lensing constraints is weaker, as the bias factor
is highly constrained due to its flat nature at low mass and hence
the improvements in bias estimates are gradual even with substantial 
improvements in mass measurements of the galaxy samples.

The situation is similar for the method~III, 
shown in Figure~\ref{fig:future}$c$, to the case in the SDSS.
As more volume is available,
the sample variance in abundance decreases even further in the future surveys,
making the uncertainties in the number density nuisance. Consequently,
the uncertainties in the clustering amplitude and the mean mass are the
dominant factors in $\Delta\rms$ at all mass range, even though the lensing 
constraints in the mass measurements improve more than the clustering 
constraints, compared to the SDSS. The best constraints on the matter
fluctuation amplitude in the future surveys
are achieved at slightly lower mass than in the SDSS, since much fewer 
galaxies at high mass are available at high redshift than at low mass,
relatively. 

All of the three methods dramatically improve the constraints on the matter 
fluctuation amplitude over the SDSS, promising sub-percent level measurements.
While the method~III has somewhat better statistical power in the SDSS
than the method~I or~II, its statistical power is comparable to other
methods in the future surveys and is {\it not}
superior to the combined method I+II in Figure~\ref{fig:future}$d$.
While no systematic errors are considered in Figure~\ref{fig:future},
they are relatively well known for the methods~I and~II.
The systematic errors we considered for the method~III 
in Sec.~\ref{ssec:sys} are relatively
innocuous, but the method~III may be subject to other unknown systematic 
errors, which may become dominant when the overall constraints reach the 
sub-percent level in precision.

\subsection{Cosmic Shear Power Spectrum from Gravitational
Lensing}
In addition to the measurements of galaxy clustering and galaxy-galaxy 
lensing, the future galaxy surveys have a large number of background galaxies 
at high redshift enough to provide robust measurements of cosmic shear signals.
Here we compute constraints on the matter fluctuation amplitude by using the
cosmic shear measurements in the future surveys and compare them to
the constraints described in Sec.~\ref{ssec:combine}.

The shapes of background source galaxies are subtly distorted by the foreground
matter distribution, and the statistical measurements of the background
galaxy shapes can be used to isolate the distortion from their intrinsic
shapes. These cosmic shear signals measured by the distortion in shapes are
represented by the projected matter fluctuation, or the convergence~$\kappa$.
Using the Limber approximation we compute the convergence power spectrum as
\beeq
C^\kappa_l=\int_0^\infty dz~W^2(z)~P_m\left(k={l\over r};z\right)~,
\eneq
where the weight function is
\beeq
W(z)={3H_0^2\over2}~{\OM\over a}\int_z^\infty dz_s~
{d\nbg\over dz}{r(z_s,z)\over r(z_s)}~.
\eneq
Here we consider different parameters for the background source distribution 
in Eq.~(\ref{eq:nbg}) that are more adequate for higher redshift surveys
than in the
SDSS: $a=2$ and $b=1.5$. The median redshifts of the source galaxies in
each survey are listed
in Table~\ref{tab:future}.
To derive the constraints from the cosmic shear measurements, we compute
the Fisher matrix \cite{HUTAET06,MAHUHU06} 
\beeq
F_{\alpha\beta}=\sum_{l=2}^{l_\up{max}}(2l+1)
{f_\up{sky}\Delta l\over2\left[C^\kappa_l+C^n_l\right]^2}
{\partial C^\kappa_l\over\partial p_\alpha}
{\partial C^\kappa_l\over\partial p_\beta}~,
\label{eq:fish}
\eneq
where the noise power spectrum in this case is
$C^n_l=\gamma_\up{int}^2/ \bar\nbg$.

A few caveats are in order, regarding our forecasts of the cosmic shear 
measurements. Equation~(\ref{eq:fish}) accounts for the galaxy shape noises as
the sole source of errors, while it is known that there exist numerous
systematic errors associated with the instruments \cite{HIMAET04,MHSGET05},
the photometric redshift measurements \cite{MAHUHU06,NAMAET12}, 
and the intrinsic alignments \cite{BLMCSE11}.
Our optimistic assumption should
serve as the best possible constraints derivable by using the cosmic shear
measurements. Furthermore, while tomographic measurements of the
cosmic shear signals at different redshift bins can improve 
constraints on the dark energy equation-of-state or its 
time evolution (e.g., \cite{HUJA04}), 
we consider only a single redshift bin for 
cosmic shear measurements in the future surveys, since 
the information contents on the matter fluctuation amplitude should be
the same as those obtained by using multiple tomographic bins.

The left panel of Figure~\ref{fig:shear} shows the redshift distributions of
background source galaxies in the future surveys. While the redshift
distribution parameters ($a,b,c$)=$(2,1.5,1)$ are different from those
in the SDSS, the shape itself (various curves)
is similar to each other, and only the
median redshifts differ for various future surveys, as listed in
Table~\ref{tab:future}. The middle panel shows the convergence power spectrum
$C_l^\kappa$ (thick) and its noise power spectrum $C_l^n$
(thin) from the cosmic shear measurements.
With few background galaxies in the SDSS, the convergence power spectrum
(solid) is dominated by the noise power spectrum, already at $l>20$,
though there exist $(2l+1)$ modes to be added per each angular multipole.
The convergence power spectra in the future surveys are larger than in the
SDSS, since the background source galaxies are at higher redshift. The noise
power spectra are smaller, simply due to larger number density in the
future surveys.

The right panel shows the $\rms$-constraints using the cosmic shear
measurements as a function of the maximum
angular multipoles. The $\rms$-constraints improve as more modes are included,
but they saturate, once the noise power spectrum overwhelms the convergence
power spectrum. In our most optimistic consideration, the cosmic shear
measurements will
provide amplitude constraints from the future surveys that are well 
below 0.1\%. 
It would appear that the cosmic
shear measurements in the future surveys trump the previous three methods
in constraining the matter fluctuation amplitude, if all the systematic
errors are under control and full statistical powers are utilized as we
assumed here. This is mostly a consequence of the fact that cosmic shear 
measurements can extract useful information on the dark matter clustering 
well into the nonlinear regime at high angular multipole~$l$. 

However, there are caveats to 
this conclusion. One is that baryonic effects also change the predictions at 
high~$l$, and these would need to be understood in detail \cite{SEHOET11}. 
Second caveat is that there exist a sample variance and a nonlinear 
contribution of the matter
trispectrum to the covariance of the lensing power spectrum
on small scales \cite{COHU01,FOMSWG,SAHAET09}. 
High degree of correlations between modes
implies no further information can be extracted from the lensing power 
spectrum on smaller scales. In the right panel of Figure~\ref{fig:shear},
the horizontal lines show these contributions to the lensing 
measurement uncertainties in the matter fluctuation amplitude
in future surveys.
We find that most of the experiments saturate the useful information,
long before they run out of signals, and the constraints from the
cosmic shear measurements are comparable to those from the combined 
method I+II.

\section{Discussion}
\label{sec:discussion}
Cosmological methods in large-scale structure are traditionally divided into 
weak lensing, galaxy clustering, and cluster abundance. While they are usually 
presented as separate techniques, there exists a 
considerable overlap between three methods. For example, the 
cluster abundance method cannot exist without a proper cluster-mass 
calibration and it is widely accepted that mass calibration 
based on weak lensing is required \cite{WEMOET12}. 
This combines cluster abundance and weak lensing. Second, a 
nuisance parameter in cluster
abundance method is the scatter between cluster observable, 
such as luminosity (in 
optical, X-ray or SZ), and the cluster mass. This scatter can in principle 
be determined from external data \cite{RORYET09,ROWEET10}, but it is not clear 
that such an approach is reliable. A more conservative approach is to determine scatter
internally from the clustering 
amplitude \cite{HU03,LIHU04,OGTA11}. 
This method thus combines lensing, clustering and counting (abundance). 
It is useful to ask the questions such as what is the mass 
range where we get most of the information, how can we combine the information from 
different mass ranges, what happens if we drop individual 
observations etc. The purpose of this paper was to address these questions. 

The Sloan Digital Sky Survey (SDSS) has enabled high precision measurements of
galaxy clustering and gravitational lensing for various galaxy samples
that cover a wide range of halo masses, so we started our analysis with the
SDSS. With careful modeling of the SDSS 
galaxy samples, we have examined the
cosmological constraining power that is contained in each galaxy sample and
can be derived by combining both measurements of galaxy clustering and 
gravitational lensing. Furthermore, we have extended our analysis 
to other ongoing and future galaxy surveys with larger volumes at higher 
redshifts such as the DES, Euclid, and the LSST.
Our joint analysis of gravitational lensing and galaxy clustering
provides a guidance to observational applications and planning future
large-scale galaxy surveys. However, in doing the analysis we had to make a 
number of simplifying assumptions, so our results should be considered as a
useful guidance rather than conclusive. 

We have investigated the cosmological
constraining power in three model-independent ways to gain physical insights
of its information contents, using constraints on the amplitude of the 
dark matter power spectrum as the figure of merit. The first method \cite{SEMAET05}
is based on the theoretical
prediction of the relation between the halo mass and the bias factor
(e.g., \cite{MOWH96,SHTO99,SMT01,SEWA04,TIROET10}).
While the large-scale clustering measurements provide only the relative
bias, galaxy-galaxy lensing measurements on small scales can provide robust
estimates of the mean mass of the galaxy samples. The theoretical 
prediction of galaxy bias can in principle be refined to arbitrary accuracy
by using numerical simulations, and its predictions depend on cosmology.
Comparison of its prediction based on the lensing measurements of mass
to the clustering measurements then determines the acceptable sets of
cosmological parameters.
This method works best at relatively high masses 
corresponding to the LRG sample or small clusters, at which the mass-dependence of the bias 
function on cosmology is strong and the uncertainties in mass 
measurements are smallest due to their
large volume coverage and high mass. Moreover, this method can be 
easily extended to the future surveys without much 
difficulty in theory and observation. 
The main observational systematic issue related to the method are the centroiding issue, which can be ameliorated if one 
uses lensing statistics that do not use small scale lensing information 
\cite{MASEET10}.
On the other hand, we have shown that this method suffers less from the overall lensing 
calibration error than the other weak lensing based methods. 
The method also assumes that the halo bias depends on the halo mass only. This is true by definition if all the halos in a 
certain mass range are included in the analysis, otherwise one must resort to halo selections based on an observable that is not 
sensitive to halo bias variations \cite{CRGAWH07}. 

The second method is to combine the large-scale measurements of galaxy-galaxy
lensing and galaxy clustering, which yields the linear theory constraints
on $\rms\OM^{0.55-1}$ \cite{YTWZKD06,BASMET10,MASLET12}. In the formulation used 
here the method has very few theoretical assumptions, since it is based on large 
scale clustering and lensing, where the two probes trace the same LSS with the 
cross-correlation coefficient close to unity \cite{BASMET10}.
While the measurement uncertainties 
depend on mass via its volume and total abundance, this method works well for 
a broad range of mass, with the best constraints in SDSS corresponding to the range between the $L_*$ galaxies to 
the low mass clusters, which allows multiple galaxy samples to be
combined to statistically tighten the resulting constraints. Furthermore,
this method can be greatly improved with larger number density of background
source galaxies at higher redshift in the future surveys, as its measurement
uncertainties lag behind the clustering measurement uncertainties in the SDSS.
In contrast, the cosmological constraint is the combination of $\rms\OM$, 
such that other constraints should be combined to break the degeneracy.
The future galaxy surveys we considered improve the constraints derived
for the SDSS, mainly by going deeper in redshift and covering larger sky. 
However, the photometric galaxy surveys 
like the DES lack capability to map galaxies in three-dimensional positions
and measure the angular clustering measurements, which become the limiting
factor given the substantial improvements in lensing measurements.

The third method is to utilize the (galaxy or cluster) abundance information, 
in conjunction with the lensing measurements of mean masses of galaxy samples. 
Due to the exponential sensitivity of the halo mass function to cosmological
parameters in the clusters range, this method can yield some of the best constraints among the three methods, 
and the mass range of the LRGs to low mass clusters works best for this method too.
The constraints are, however, comparable to the combined method I+II,
which use both large-scale and small-scale lensing information, 
in addition to the large-scale clustering information. 
While the abundance is modeled by assuming a log-normal scatter in the
mass-observable relation, our investigation of the systematic errors in this 
method shows that the systematic errors such as the presence $\alpha_g$
of invisible halos and the skewness $S_3$ of the probability distribution 
in the mass-observable relation are relatively innocuous given the present 
level of uncertainties in those parameters.
For the future surveys this method continues to yield competitive constraints, 
but is not superior to  the combined method I+II.
The methods~I and~II 
do not use abundance information and are thus not sensitive to 
the systematics associated with that, such as identifying multiple clusters in the 
same halo.

We also compared the derived constraints to those from cosmic shear measurements that 
use shear-shear correlations. These can use information to significantly smaller scales 
and as a result can provide tighter constraints, despite the
two-dimensional analysis assumed here. 
However, we argue that some of the gains cannot be achieved because of the baryonic effects \cite{SEHOET11},
and more importantly, the sample variance and the
nonlinear evolution of the matter trispectrum make
the lensing power spectra on small scales highly correlated 
\cite{COHU01,FOMSWG,SAHAET09},
setting the lower limit to the figure of merit one can obtain from the lensing
power spectrum. Consequently, the constraints from the cosmic shear 
measurements are as strong as that from the combined method I+II.
Furthermore,
shear-shear analysis may also be more sensitive to the various spurious systematics such as variable PSF. 
Considering the serious systematic uncertainties present in the cosmic shear method it is 
useful to consider alternative methods, and the three
methods explored in this paper are the 
best alternatives proposed so far in measuring perturbations of dark matter, 
apart from redshift-space distortions which we do not consider in this paper. Their combined power 
is comparable to  that of the shear-shear 
power spectrum, with very different systematics errors. In particular, the combined 
lensing and clustering analysis method, which combines the methods~I and~II 
of this paper, 
is a method that has not been discussed much in the literature, yet it 
gives predicted errors for SDSS that are comparable to 
the better known cluster abundance method with weak lensing calibration
(method~III). 
The method is currently being applied to the SDSS data and we expect this method will play 
an important role in the future surveys as well. 

\acknowledgments
We acknowledge useful discussions with Pat McDonald and Rachel Mandelbaum.
We thank Chris Hirata for providing us with trispectrum contribution to
the cosmic shear measurements.
This work is supported by the Swiss National Foundation (SNF) under contract
200021-116696/1 and WCU grant R32-10130.
J.Y. is supported by the SNF Ambizione Grant.

\bibliography{ms.bbl}

\vfill

\end{document}